\documentclass[fleqn,usenatbib]{mnras}

\usepackage{newtxtext,newtxmath}

\usepackage[T1]{fontenc}

\DeclareRobustCommand{\VAN}[3]{#2}
\let\VANthebibliography\thebibliography
\def\thebibliography{\DeclareRobustCommand{\VAN}[3]{##3}\VANthebibliography}

\usepackage{graphicx}	
\usepackage{amsmath}	
\usepackage{soul}
\usepackage{rotating}

\defcitealias{brout_its_2021}{BS21}

\title[SN Ia Hubble Residual vs. Host \ion{O}{II}]{[\ion{O}{II}] as an Effective Indicator of the Dependence Between the Standardised Luminosities of Type Ia Supernovae and the Properties of their Host Galaxies}

\author[B. Martin et al.]{
B. Martin$^{1}$\thanks{E-mail: bailey.martin@anu.edu.au},
C. Lidman$^{1,2}$,
D. Brout$^{3,4}$,
B. E. Tucker$^{1,5,6}$,
M. Dixon$^{7}$,
P. Armstrong$^{1}$
\\
$^{1}$The Research School of Astronomy and Astrophysics, The Australian National University, ACT 2601, Australia\\
$^{2}$Centre for Gravitational Astrophysics, College of Science, The Australian National University, ACT 2601, Australia\\
$^{3}$Department of Astronomy, Boston University, 725 Commonwealth Ave., Boston, MA 02215, USA\\
$^{4}$Department of Physics, Boston University, 590 Commonwealth Ave., Boston, MA 02215, USA\\
$^{5}$National Centre for the Public Awareness of Science, The Australian National University, ACT 2601, Australia\\
$^{6}$The ARC Centre of Excellence for All-Sky Astrophysics in 3 Dimensions (ASTRO 3D), Australia\\
$^{7}$Centre for Astrophysics \& Supercomputing, Swinburne University of Technology, Victoria 3122, Australia
}

\date{Accepted XXX. Received YYY; in original form ZZZ}

\pubyear{2024}

\begin{document}
\label{firstpage}
\pagerange{\pageref{firstpage}--\pageref{lastpage}}
\maketitle

\begin{abstract}
%
We have obtained IFU spectra of 75 SN Ia host galaxies from the Foundation Supernova survey to search for correlations between the properties of individual galaxies and SN Hubble residuals. After standard corrections for light-curve width and SN colour have been applied, we find correlations between Hubble residuals and the equivalent width of the [\ion{O}{II}] $\lambda\lambda$ 3727, 3729 doublet (2.3$\sigma$), an indicator of the specific star formation rate (sSFR). When splitting our sample by SN colour, we find no colour dependence impacting the correlation between EW[\ion{O}{II}] and Hubble residual. However, when splitting by colour, we reveal a correlation between the Hubble residuals of blue SNe Ia and the Balmer decrement (2.2$\sigma$), an indicator of dust attenuation. These correlations remain after applying a mass-step correction, suggesting that the mass-step correction does not fully account for the limitations of the colour correction used to standardise SNe Ia. Rather than a mass correction, we apply a correction to SNe from star forming galaxies based on their measurable EW[\ion{O}{II}]. We find that this correction also removes the host galaxy mass step, while also greatly reducing the significance of the correlation with the Balmer decrement for blue SNe Ia. We find that correcting for EW[\ion{O}{II}], in addition to or in place of the mass-step, may further reduce the scatter in the Hubble diagram.

\end{abstract}




\section{Introduction}
Accurate distance measurements are the cornerstone of observational cosmology. One of the most important and widely used distance indicators are Type Ia supernovae (SNe Ia). Their use as distance indicators led to the discovery of the accelerated expansion of our Universe \citep[][]{riess_observational_1998,perlmutter_measurements_1999}, and have since been used to place tighter and tighter constraints on cosmological parameters that describe the  expansion of our Universe, such as the dark energy equation of state and the Hubble constant \citep[e.g.][]{betoule_improved_2014, scolnic_complete_2018, brout_first_2019, khetan_new_2021,freedman_measurements_2021,riess_comprehensive_2022,rubin_union_2023, vincenzi2024dark, des_collaboration_dark_2024}.\par 

Through observations, so-called `Branch-normal' SNe Ia have been found to reach a peak B-band absolute magnitude of roughly $M_B \sim -19.2$ \citep{riess_comprehensive_2022}, with an rms scatter of $\sim 0.5$ mag. By applying corrections derived from multi-colour SN Ia light-curves, the observed scatter in peak brightness can be significantly reduced to about 0.15 mag, and therefore SNe Ia are described as standardisable candles.\par

The most common corrections applied to the distance modulus are the SN Ia colour \citep{tripp_two-parameter_1998} and the light-curve width corrections \citep{phillips_absolute_1993}, both of which are correlated with the peak magnitude of the SN Ia. However, after the standard light-curve corrections are applied, we are still left with a scatter that is larger than measurement uncertainties, meaning that there is a level of intrinsic scatter that may be reduced with additional data. The physical nature of this scatter is unknown.\par

After the standard light-curve corrections are applied to SNe Ia, it has been shown that more massive SN host galaxies produce SNe Ia that are systematically brighter than those from low mass galaxies \citep{sullivan_dependence_2010, kelly_hubble_2010, childress_host_2013, uddin_cosmological_2017, smith_first_2020,kelsey_effect_2021,dixon_using_2022}. This is modelled as a step function centreed at 10$^{10}\ M_\odot$, and is referred to in the literature as the `mass-step'.\par

A leading candidate for the origin of the mass-step is our treatment of dust attenuation. \citet{salim_dust_2018} find that high-mass galaxies tend to have shallower attenuation laws overall, but show a greater amount of attenuation in the V-band compared to low-mass galaxies. By allowing the amount of dust attenuation and the attenuation law to vary as a function of host galaxy mass, the mass-step can be reproduced as a function of SN colour \citep{brout_its_2021,meldorf_dark_2022,popovic_pantheon_2022}. \citet{brout_its_2021} are also able to reproduce the trend and RMS scatter in Hubble residual as a function of SN colour and host stellar mass.\par

Other host galaxy properties have been shown to correlate with SN Ia luminosities after standardisation. These include host galaxy metallicity \citep{pan_host_2014, campbell_how_2016, galbany_aperture-corrected_2022}, specific star formation rate \citep{lampeitl_effect_2010, childress_host_2013, rigault_strong_2020, galbany_aperture-corrected_2022, dixon_using_2022} and the equivalent width of \ion{H}{$\alpha$} \citep{galbany_aperture-corrected_2022}. \par

In this paper, we use IFU spectra to measure a broad range of SN Ia host galaxy properties, and identify correlations between these properties and the SN Hubble residuals, once SN Ia luminosities have been corrected for colour and light-curve width. We obtain spectroscopic properties of our host galaxies, rather than the more extensively studied photometric properties \citep[e.g.][]{sullivan_dependence_2010,rigault_evidence_2013,uddin_cosmological_2017,rose_think_2019,smith_first_2020,kelsey_effect_2021}, as spectra provide more detailed information than broad-band photometry.\par

\begin{figure}
    \centering
    \includegraphics[width = \linewidth]{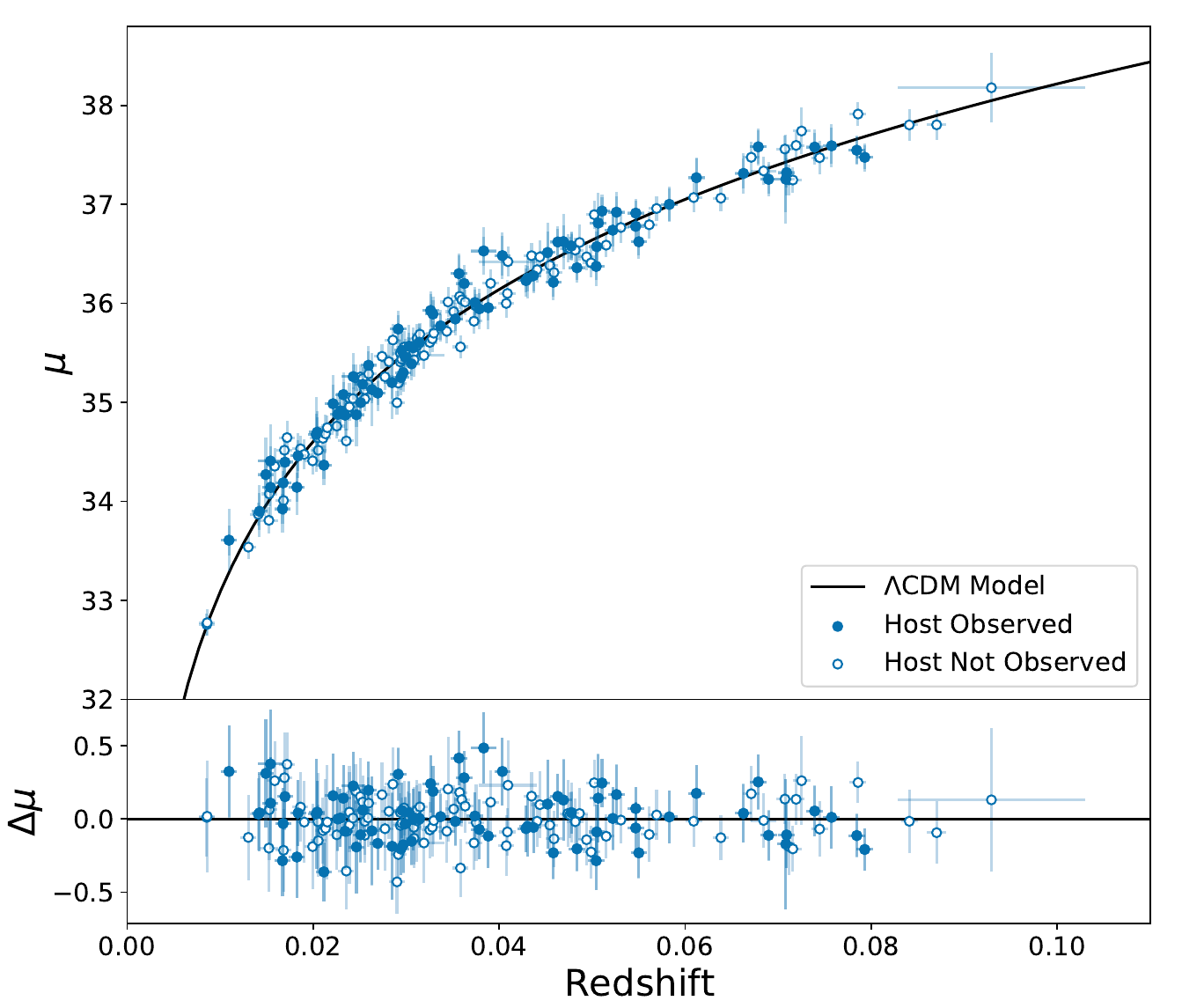}
    \caption{Above: A Hubble Diagram showing the light-curve width and colour corrected distance moduli of cosmologically-useful SNe Ia from the Foundation Survey. Those SNe whose host spectra we have obtained are shown by the filled data points. The predicted relationship for a Flat-$\Lambda$CDM cosmology ($w = -1,\ \Omega_m = 0.326,\ H_0 = 73.04$ km s$^{-1}$ Mpc$^{-1},\ M_B = -19.253$) is shown by the black curve. Below: The `Hubble residuals' of the Foundation SNe, relative to this cosmological model.}
    \label{fig:Hubble Diagram}
\end{figure}

Previous spectroscopic studies of SN Ia host galaxies include \citet{childress_host_2013}, who obtained long-slit spectra of individual host galaxies to look for relationships between the SN Ia distance measurement and the gas-phase metallicity of the host. \citet{dixon_using_2022} stacked host galaxy spectra together to obtain a high enough signal-to-noise Ratio (S/N) to measure properties of their hosts, turning their sample of 625 host galaxies into 12 data points which they could use for their correlation analysis. In our work, we obtain IFU spectra of 75 galaxies that have hosted SNe Ia from the Foundation survey \citep{foley_foundation_2018}. These spectra have a sufficient S/N for us to measure the properties of the host galaxies individually. \par

This paper is composed of the following sections: In Section \ref{sec:Data} we describe the sample of SNe Ia and their hosts used in this analysis.  The methods we used to derive their spectral properties are outlined in Section \ref{sec:Methods}. We present the results of our analysis in Section \ref{sec:Results}, followed by a discussion of these results and their implications in Section \ref{sec:Discussion}. Finally, we provide our concluding remarks in Section \ref{sec:Conclusion}.

\section{Data Selection} \label{sec:Data}

\subsection{The Foundation Survey}

In our analysis, we utilise the sample of SNe Ia from the Foundation Supernova Survey \citep{foley_foundation_2018, jones_foundation_2019}. Foundation observed a large low-redshift ($z<0.1$) sample of SNe Ia using the Pan-STARRS telescope, providing a low redshift anchor for the Hubble diagram. As Foundation is a low-redshift sample, their hosts can be observed to a moderately high S/N with hour-long integration times on 2-metre class telescopes. This makes Foundation an ideal sample of SNe Ia for host galaxy spectroscopic analyses.

\subsection{Hubble Residuals}

\begin{figure}
    \centering
    \includegraphics[width = \linewidth]{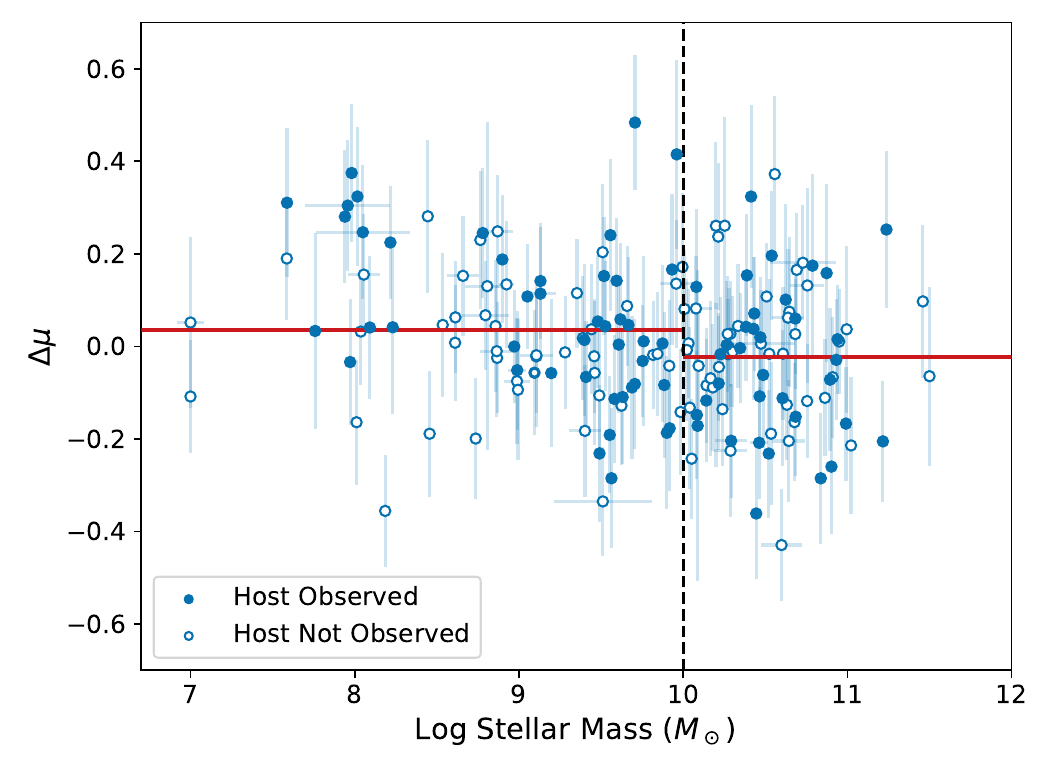}
    \caption{The Hubble residuals of Foundation SNe Ia against host galaxy stellar mass. Those SNe whose host spectra we have obtained are shown by the filled data points. By computing the mean Hubble residuals on either side of the canonical mass cut at $10^{10}M_\odot$ (black dashed line), we obtain a mass-step of $0.10 \pm 0.03$ for our observed sample, in comparison to $0.06 \pm 0.02$ for the full Foundation sample (red line).}
    \label{fig:Mass Step}
\end{figure}

\begin{figure*}
    \centering
    \includegraphics[width = \linewidth]{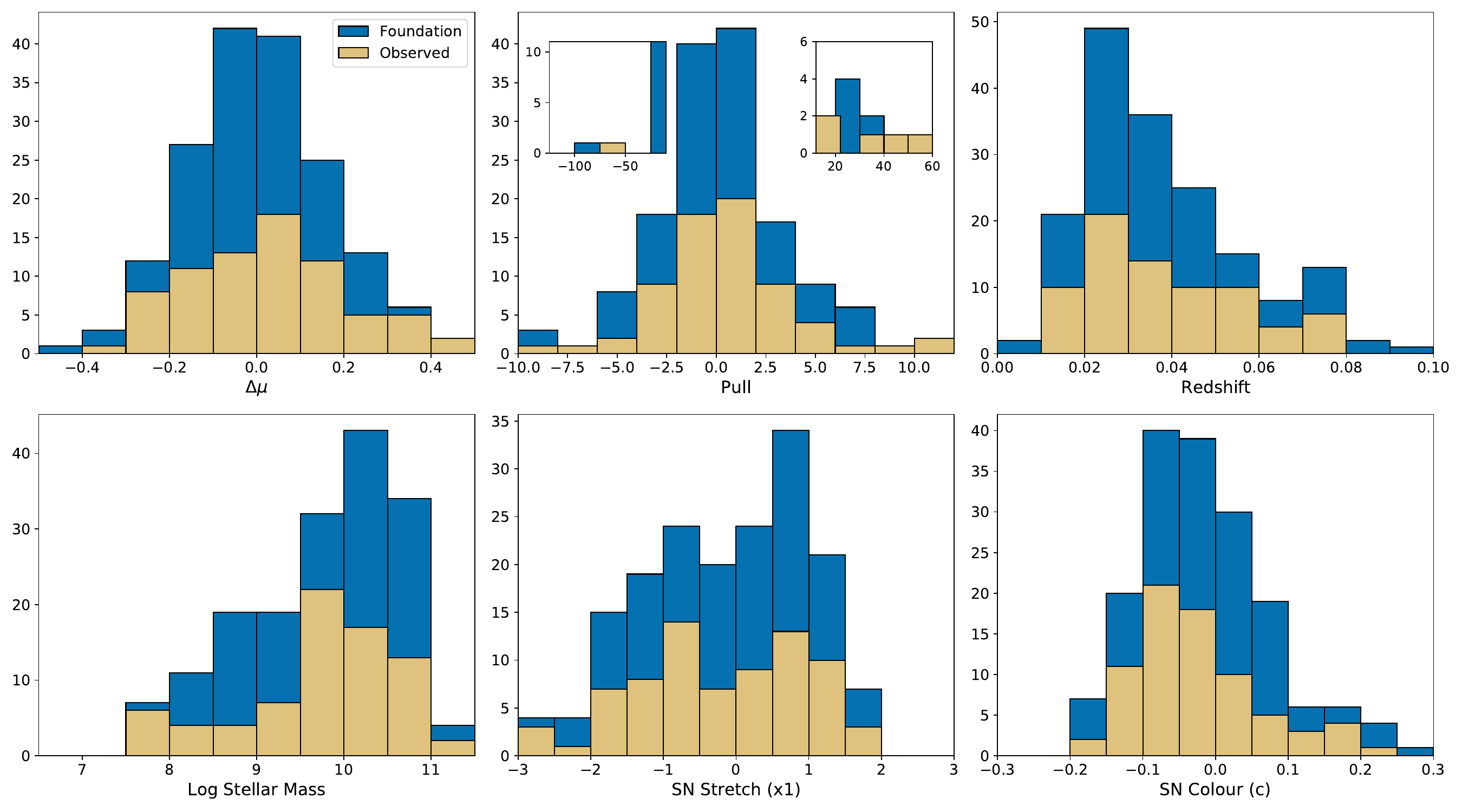}
    \caption{The distributions of Hubble residuals, pulls, redshifts, host galaxy stellar masses, and SN Ia light-curve parameters of our observed galaxy sample, compared with the distributions for all Foundation SNe. Our subsample of 75 SN Ia host galaxies adequately covers the range of SN and host parameters seen in the full Foundation sample.}
    \label{fig:Sample Hists}
\end{figure*}

The Hubble residual ($\Delta\mu$) is the offset between the measured distance and the predicted distance from a cosmological model, at a given redshift: 
\begin{equation}
    \Delta\mu = \mu_\mathrm{obs} - \mu_\mathrm{model}(z)
\end{equation}

\noindent We utilise existing Foundation SN light-curve fits from the Pantheon+ analysis \citep{brout_pantheon_2022}. These light-curve fits have been derived using the SALT2 algorithm \citep{guy_salt2_2007, guy_supernova_2010,brout_SuperCal_2022}, which provides standard light-curve parameters used in the following distance modulus parameterisation \citep{tripp_two-parameter_1998},
\begin{equation}\label{eq:Tripp}
    \mu_\mathrm{obs} = m_B - M_B + \alpha x_1 - \beta c
\end{equation}
where $m_B$ is the peak observed B-band magnitude of the SN, $x_1$ is a measure of the light-curve width, and $c$ represents the SN colour. $M_B$ represents the peak absolute magnitude of a SN Ia, which we set to $M_B = -19.253$ mag based on the fit to the Pantheon+SH0ES dataset \citep{riess_comprehensive_2022}, while $\alpha$ and $\beta$ are global nuisance parameters relating the light-curve width and colour to SN luminosity. We take the values of $\alpha = 0.148$ and $\beta = 3.09$ derived by the Pantheon+ analysis \citep{brout_pantheon_2022} for an intrinsic scatter model developed by \citet{brout_its_2021}.\par

It is customary to include corrections to SN Ia distance moduli for selection biases, and for the host galaxy mass step. These corrections are provided by the Pantheon+ analysis \citep{brout_pantheon_2022}, however we omit these corrections in our analysis as we seek to identify correlations with other host properties. In Section \ref{sec:Applying Mass Step} we apply these bias and host mass corrections to analyse their impact on our results. \par

We use the Pantheon+ flat $\Lambda$CDM cosmological model \citep[$w = -1,\ \Omega_m = 0.326,\ H_0 = 73.04$ km s$^{-1}$ Mpc$^{-1}$,][]{brout_pantheon_2022} when deriving model distance moduli, allowing us to obtain Hubble residuals for the Foundation SNe, as shown in Figure \ref{fig:Hubble Diagram}. From Figure \ref{fig:Hubble Diagram}, we find that after the standard light-curve corrections for stretch and colour, the Foundation SNe have a scatter of $0.16 \pm 0.02$ mag.\par

By comparing SN Ia Hubble residuals, which have been corrected for light-curve width and colour, to properties of their host galaxies, we can identify and characterise any systematic influences galaxies have on the brightness of their SNe . The well-established presence of the `mass-step,' by which brighter standardised SNe Ia are observed in galaxies with a stellar mass above $10^{10}M_\odot$, is an example of such an influence. The mass-step for the Foundation dataset is shown in Figure \ref{fig:Mass Step}, demonstrating that SNe Ia from high-mass galaxies are on average $0.06\pm0.02$ mag brighter than SNe Ia from low-mass galaxies.

\subsection{Host Galaxy Sample}

\begin{figure*}
    \centering
    \includegraphics[width = \linewidth]{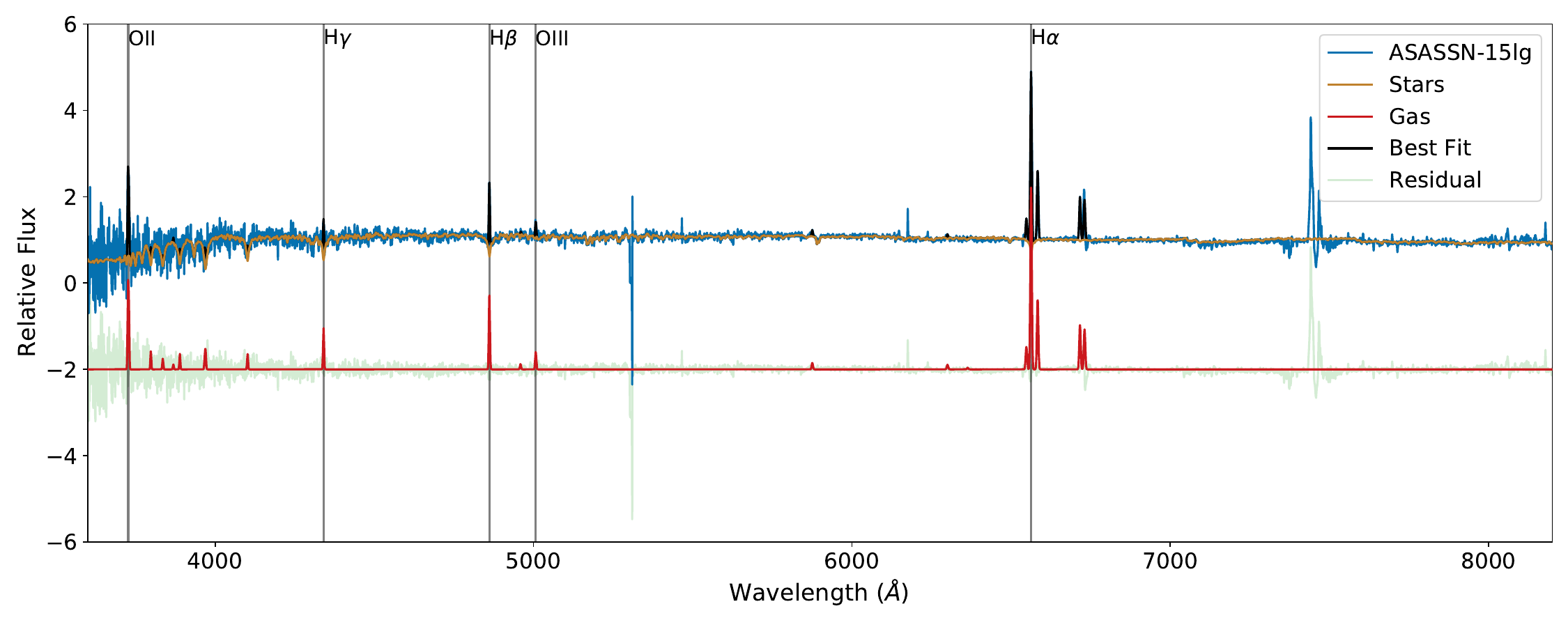}
    \includegraphics[width = \linewidth]{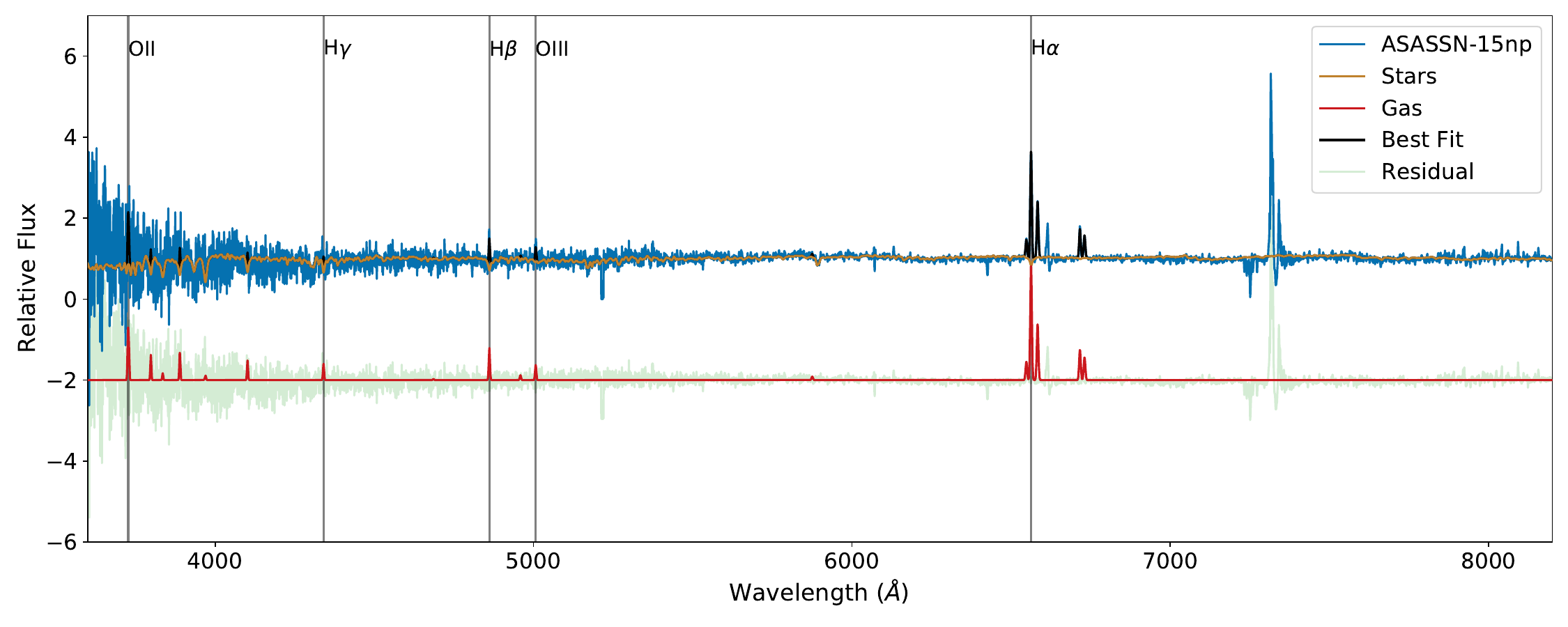}
    \includegraphics[width = \linewidth]{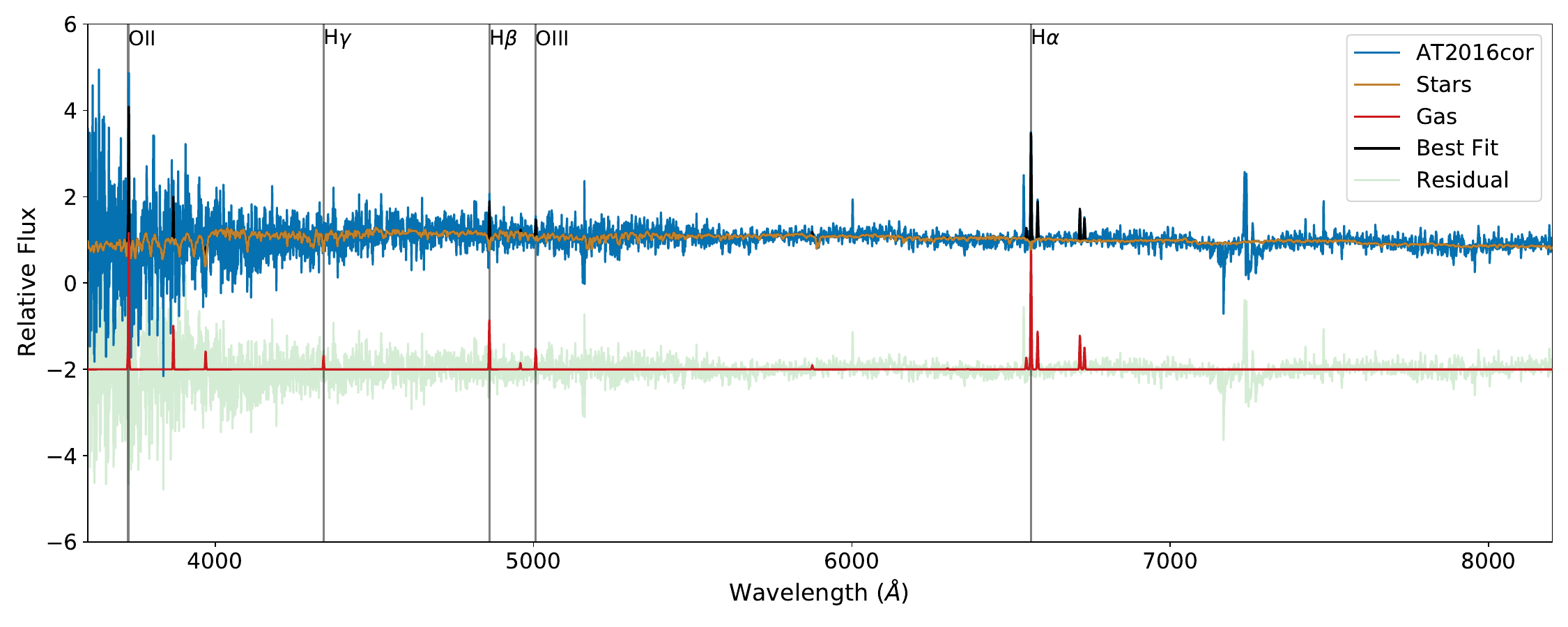}
    \caption{pPXF fits to the observed host galaxies of ASASSN-15lg, ASASSN-15np and AT2016cor, with median signal-to-noise ratios of 17, 11 and 9, respectively. The best-fitting spectrum is shown in black. This is split into stellar and gaseous components, shown in orange and red, respectively. The terrestrial A-band feature ($7600$\AA) produces a large residual at $\sim 7200-7500$\AA, due to our spectra being plotted in the rest-frame of the host galaxy.}
    \label{fig:Example Fit}
\end{figure*}

We observed a sample of 75 targets from the full list of 179 cosmologically-useful SNe Ia from the Foundation Supernova Survey. The sample was chosen based on a number of criteria.\par

Firstly, we removed all targets at a declination $\delta > 28^\circ$, as the host galaxies do not rise above an airmass of 2 when viewed from Siding Spring Observatory (SSO) in Australia. This reduced the size of our potential host galaxy sample from 179 to 131. \par

We then prioritised observations of host galaxies whose SNe have the most extreme, but also most accurate Hubble residual measurements. We quantified this by determining the `pull' parameter for each SN Ia, which is given by:

\begin{equation}\label{eq:Pull}
    \mathrm{Pull} = \frac{|\Delta\mu|}{\sigma_\mu}
\end{equation}

\noindent Here $\Delta\mu$ is the Hubble residual, and $\sigma_\mu$ is the uncertainty in the measurement of the distance modulus $\mu$. We computed the pull parameter for each SN Ia using their $m_B$, $x_1$ and $c$-corrected distance moduli. We preferentially selected galaxies with the highest pull values, which arise from either large Hubble residuals or precise measurements of the distance modulus $\mu$.\par

\begin{table*}
    \centering
    \caption{For each measured host galaxy property, we list the number of of hosts with a reliable measurement of this property, along with the slope and significance of the linear best fit when plotted against Hubble residual (Figure \ref{fig:Included Properties}), the intrinsic scatter about this fit, and the Pearson and Spearman correlation coefficients. The most significant trends come from the equivalent widths of the strong gas emission lines, many of which correlate with specific star formation rate.}
    \begin{tabular}{l c c c c c c}
    \hline
    Host Property & No. of Hosts & Slope & Significance $(\sigma)$ & Intrinsic Scatter & Pearson $r$ & Spearman $r$\\
    \hline
    Mass-to-Light Ratio & $75$ & $-0.004\pm 0.014$ & $0.26$ & $0.05\pm 0.03$ & $0.03\pm 0.09$ & $-0.08 \pm 0.09$\\
    Log Stellar Age & $75$ & $-0.027\pm 0.041$ & $0.66$ & $0.05\pm 0.03$ & $-0.10\pm 0.09$ & $-0.08 \pm 0.09$\\
    Stellar Metallicity [M/H] & $75$ & $-0.051\pm 0.049$ & $1.03$ & $0.05\pm 0.03$ & $-0.10\pm 0.09$ & $-0.16 \pm 0.09$\\
    Log Stellar Mass & $75$ & $-0.072\pm 0.027$ & $2.66$ & $0.04\pm 0.03$ & $-0.41\pm 0.10$ & $-0.35 \pm 0.08$\\
    $H\alpha/H\beta$ & $44$ & $0.135\pm 0.077$ & $1.75$ & $0.04\pm 0.03$ & $0.33\pm 0.14$ & $0.31 \pm 0.14$\\
    $H\gamma/H\beta$ & $37$ & $-0.153\pm 0.589$ & $0.26$ & $0.05\pm 0.04$ & $0.08\pm 0.16$ & $-0.11 \pm 0.16$\\
    EW[\ion{H}{$\alpha$}] & $63$ & $-0.006\pm 0.003$ & $2.08$ & $0.04\pm 0.03$ & $-0.34\pm 0.11$ & $-0.36 \pm 0.11$\\
    EW[\ion{H}{$\beta$}] & $47$ & $-0.020\pm 0.008$ & $2.50$ & $0.04\pm 0.03$ & $-0.51\pm 0.14$ & $-0.49 \pm 0.13$\\
    EW[\ion{H}{$\gamma$}] & $39$ & $-0.026\pm 0.012$ & $2.16$ & $0.04\pm 0.03$ & $-0.44\pm 0.14$ & $-0.44 \pm 0.15$\\
    EW[\ion{O}{I}] $\lambda 6300$ & $16$ & $-0.010\pm 0.162$ & $0.06$ & $0.06\pm 0.06$ & $-0.29\pm 0.25$ & $-0.16 \pm 0.25$\\
    EW[\ion{O}{II}] $\lambda\lambda 3726,3729$ & $37$ & $-0.014\pm 0.006$ & $2.30$ & $0.04\pm 0.03$ & $-0.58\pm 0.15$ & $-0.58 \pm 0.14$\\
    EW[\ion{O}{III}] $\lambda 5007$ & $54$ & $-0.013\pm 0.006$ & $2.27$ & $0.04\pm 0.03$ & $-0.40\pm 0.13$ & $-0.37 \pm 0.13$\\
    EW[\ion{S}{II}] $\lambda\lambda 6716,6731$ & $55$ & $-0.018\pm 0.009$ & $2.03$ & $0.04\pm 0.03$ & $-0.46\pm 0.13$ & $-0.29 \pm 0.12$\\
    EW[\ion{N}{II}] $\lambda 6583$ & $61$ & $-0.014\pm 0.006$ & $2.49$ & $0.04\pm 0.03$ & $-0.41\pm 0.10$ & $-0.33 \pm 0.11$\\
    EW[\ion{Ne}{III}] $\lambda 3869$ & $15$ & $0.083\pm 0.066$ & $1.26$ & $0.08\pm 0.07$ & $0.39\pm 0.20$ & $0.24 \pm 0.22$\\
    EW[\ion{He}{I}] $\lambda 5876$ & $25$ & $-0.176\pm 0.164$ & $1.07$ & $0.05\pm 0.05$ & $-0.32\pm 0.19$ & $-0.57 \pm 0.20$\\
    \hline
    \end{tabular}
    \label{tab:correlation_stats}
\end{table*}

\begin{figure*}
    \centering
    \includegraphics[width = 0.92\linewidth]{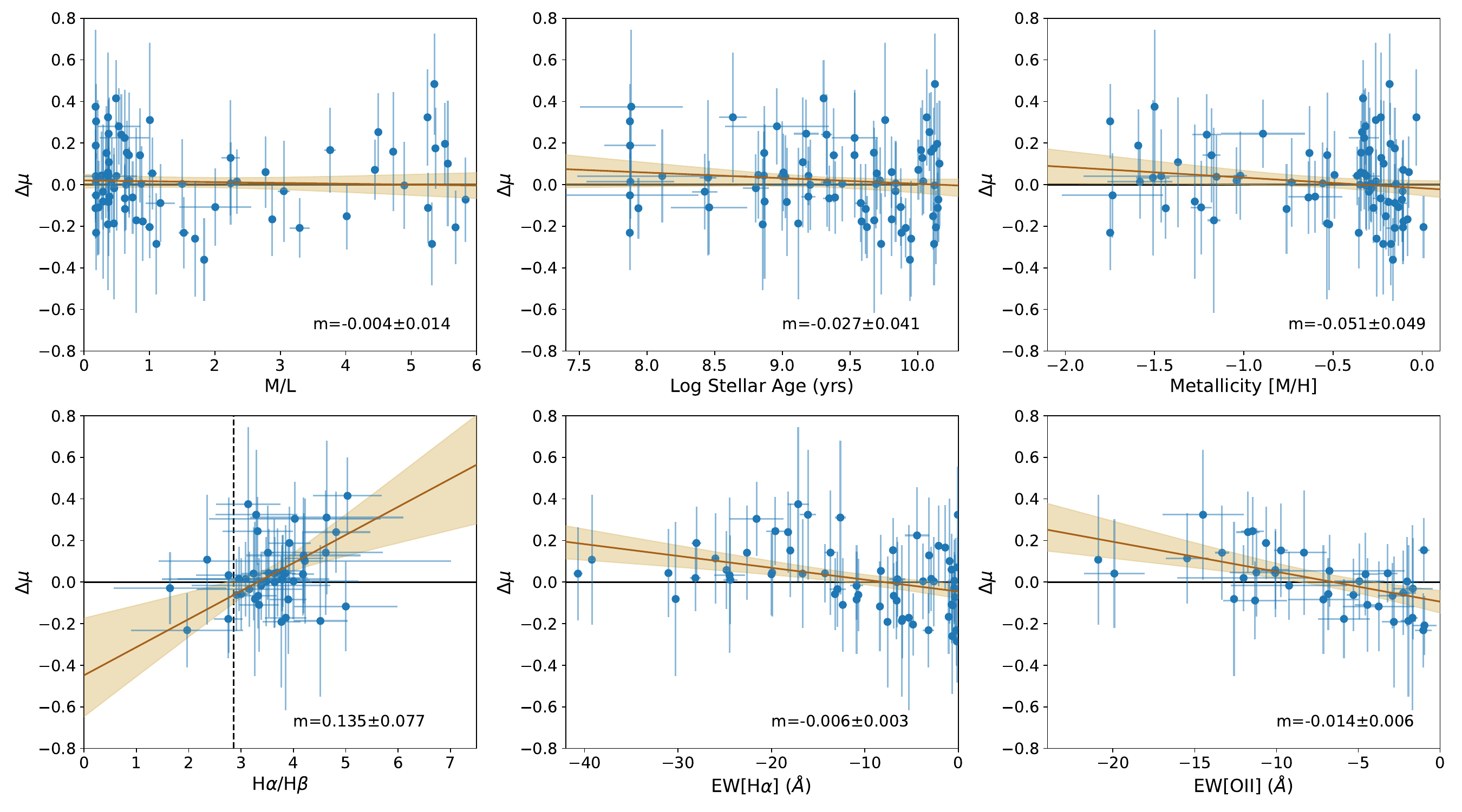}
    \caption{Fitted trends between spectroscopically-derived SN Ia host galaxy properties and their Hubble Residuals (corrected for $x_1$ and $c$). The gold line represents the best-fitting linear relationship, with the shaded region representing a 1$\sigma$ confidence interval. The slope $m$ of the fitted relationship, and its corresponding uncertainty, is given in each plot. We find the strongest trends with the \ion{H}{$\alpha$} and [\ion{O}{II}] $\lambda\lambda$ 3727, 3729 equivalent widths. The black dashed line in the lower left plot represents the theoretical lower limit of the Balmer decrement for Case-B recombination \citep{osterbrock_astrophysics_1989}. Galaxies reddened by dust lie to the right of this line.}
    \label{fig:Included Properties}
\end{figure*}

When selecting solely on this criterion, we found that our sample was heavily weighted towards high-mass galaxies. We therefore selected all observable galaxies with $\mathrm{log}(M_*) < 10$, and only applied our `pull' parameter selection to those galaxies above this mass. Based on these selection criteria and our available observing allocation, we were able to obtain spectra of 75 SN Ia host galaxies. The Hubble residual, pull, redshift, stellar mass, SN Ia light-curve width and colour distributions of our observed sample are shown in Figure \ref{fig:Sample Hists}, alongside the distributions for the full Foundation sample.

\section{Methods} \label{sec:Methods}
\begin{table*}
    \centering
    \caption{The impact of splitting the SN Ia sample into red ($c>-0.025$) and blue ($c\leq-0.025$) SNe on the relationship between Hubble residual and host stellar mass. We list the slope and significance of the best-fitting linear relationships, and the Pearson and Spearman correlation coefficients for each subset, along with the results for the combined sample.}
    \begin{tabular}{c c c c c c}
    \hline
    Colour & No. of Hosts & Slope & Significance $(\sigma)$ & Pearson $r$ & Spearman $r$\\
    \hline
    Red & $33$ & $-0.097\pm 0.051$ & $1.89$ & $-0.39 \pm 0.17$ & $-0.28 \pm 0.14$ \\
    Blue & $42$ & $-0.063\pm 0.035$ & $1.81$ & $-0.37 \pm 0.12$ & $-0.28 \pm 0.12$ \\
    Combined & $75$ & $-0.072 \pm 0.027$ & $2.66$ & $-0.41 \pm 0.10$ & $-0.35 \pm 0.08$\\
    \hline
    \end{tabular}
    
    \label{tab:mass_vs_mu}
\end{table*}
\subsection{Observations}

To measure the spectroscopic properties of SN Ia host galaxies, we utilised the Wide Field Spectrograph \citep[WiFeS;][]{dopita_wide_2007}, mounted on the ANU 2.3-metre Telescope at Siding Spring Observatory. WiFeS has a field of view of $25\times38$ arcseconds, divided into 25 $1\times 38$ arcsecond slitlets, which are projected on to the $4096\times4096$ pixel CCDs in the red and blue arms of the instrument. With $2\times$ binning in the spatial direction, we obtain spaxels of $1.0\times1.0$ arcseconds.\par

We used the R3000 and B3000 gratings, each with a resolution of $R=3000$, in combination with the RT560 dichroic, to obtain spectra covering a wavelength range of $3500 - 9000$\AA. We observed our chosen host galaxies using the Nod \& Shuffle mode, with three 1200s exposures (600s on target, 600s on sky). The median SNR across our sample of host galaxies was 15.

\subsection{Deriving Host Galaxy Properties} \label{sec:pPXF Fitting}

To extract the spectral properties of our observed host galaxies, we utilise pPXF \citep[Penalised Pixel Fitting;][]{cappellari_parametric_2004}. pPXF performs a fit to a galaxy spectrum by optimising a linear combination of single stellar population (SSP) models. Each SSP has a given age and metallicity, and the specific combinations and weightings of each SSP included in the pPXF fit allows us to estimate the mean stellar properties.\par

In our pPXF fits, we utilise stellar spectra from the Extended-MILES \citep[E-MILES][]{vazdekis_uv-extended_2016} spectral library. E-MILES is an extension to the widely used MILES \citep{sanchez-blazquez_medium-resolution_2006} catalogue, and pushes further into the UV and IR. E-MILES covers a spectral range of $1680-50000$\AA, and by assuming a \citet{salpeter_luminosity_1955} IMF, pPXF builds a best-fitting synthetic spectrum from the SSPs, with population ages ranging from $0.063-17.78$ Gyr and metallicities ranging from $-1.71$ to $+0.22\ [M/H]$.\par

The best-fitting spectrum can be broken down into stellar and gaseous components. This allows us to derive a weighted age, metallicity and Mass-to-Light ratio (M/L) for each galaxy, as well as fluxes and equivalent widths of spectral lines. Example pPXF fits to three of our observed Foundation host galaxies are shown in Figure \ref{fig:Example Fit}, representing the range of S/N ratios in our data. In order to derive uncertainties on our measured properties, we vary the flux at each wavelength bin across the spectrum in a Gaussian manner, based on the measured variance, to produce an adjusted spectrum. We create 1000 such spectra for a given host galaxy, and perform a pPXF fit to derive host properties for each of these spectra. The quoted errors on a host galaxy's properties are the standard deviations of the values computed from the adjusted spectra.\par

\section{Results} \label{sec:Results}

\subsection{Host galaxy properties vs. Hubble Residual}

We now present the results of our analysis, by comparing SN Ia Hubble residuals to the spectroscopically-derived properties of their host galaxies. We apply two S/N cuts, requiring the median S/N of each spectrum to be $>5$, and the S/N for any given emission line flux fitted by pPXF to be greater than 1. This means that the number of data points in each correlation plot varies, as they will contain a subset of our 75 host galaxy observations.\par

In Figure \ref{fig:Included Properties} we show a sample of measured properties of our SN Ia host galaxies, plotted against the Hubble residuals, which are computed in accordance with with Equation \ref{eq:Tripp}. The uncertainties in the host properties are determined with the resampling approach outlined in Section \ref{sec:pPXF Fitting}. In these plots, we show the best-fitting linear relationship, which is computed by linear regression with a hierarchical Bayesian approach using the python package LINMIX \citep{kelly_aspects_2007}. The shaded region represents the 1$\sigma$ uncertainty in this linear fit. We plot the equivalent widths of \ion{H}{$\alpha$} and the [\ion{O}{II}] $\lambda\lambda$ 3727, 3729 doublet as a representation of the relationships present for most of the emission lines. The correlation statistics for the full set of emission lines, along with the stellar properties of the host galaxies, are listed in Table \ref{tab:correlation_stats}.\par

We find in Table \ref{tab:correlation_stats} that the strongest linear fit with Hubble residual comes from the host stellar mass, which we expect due to the existence of the host galaxy mass step. The slope of the linear fit has a significance of $2.7\sigma$. However, the Pearson correlation coefficient between host stellar mass and Hubble residual is not the strongest of all the host properties, with a value of $r=-0.41\pm 0.10$.

\begin{table*}
    \centering
    \caption{The impact of splitting the SN Ia sample into red ($c>-0.025$) and blue ($c\leq-0.025$) SNe on the relationship between Hubble residual and EW[\ion{O}{II}]. We list the slope and significance of the best-fitting linear relationships, and the Pearson and Spearman correlation coefficients for each subset, along with the results for the combined sample.}
    \begin{tabular}{c c c c c c}
    \hline
    Colour & No. of Hosts & Slope & Significance $(\sigma)$ & Pearson $r$ & Spearman $r$\\
    \hline
    Red & $18$ & $-0.014\pm 0.009$ & $1.61$ & $-0.70 \pm 0.23$ & $-0.68 \pm 0.21$\\
    Blue & $19$ & $-0.012\pm 0.010$ & $1.24$ & $-0.45 \pm 0.20$ & $-0.43 \pm 0.20$\\
    Combined & $37$ & $-0.014 \pm 0.006$ & $2.30$ & $-0.58 \pm 0.15$ & $-0.58 \pm 0.14$\\
    \hline
    \end{tabular}
    
    \label{tab:OII_vs_mu}
\end{table*}

\begin{table*}
    \centering
    \caption{The impact of splitting the SN Ia sample into red ($c>-0.025$) and blue ($c\leq-0.025$) SNe on the relationship between Hubble residual and the Balmer decrement. We list the slope and significance of the best-fitting linear relationships, and the Pearson and Spearman correlation coefficients for each subset, along with the results for the combined sample.}
    \begin{tabular}{c c c c c c}
    \hline
    Colour & No. of Hosts & Slope & Significance $(\sigma)$ & Pearson $r$ & Spearman $r$\\
    \hline
    Red & $21$ & $0.002\pm 0.117$ & $0.02$ & $0.01 \pm 0.21$ & $0.05 \pm 0.21$\\
    Blue & $23$ & $0.196\pm 0.090$ & $2.19$ & $0.72 \pm 0.19$ & $0.61 \pm 0.18$\\
    Combined & $44$ & $0.135\pm 0.077$ & $1.75$ & $0.33 \pm 0.14$ & $0.31 \pm 0.14$\\
    \hline
    \end{tabular}
    
    \label{tab:Balmer_vs_mu}
\end{table*}

\begin{figure}
    \centering
    \includegraphics[width = 0.85\linewidth]{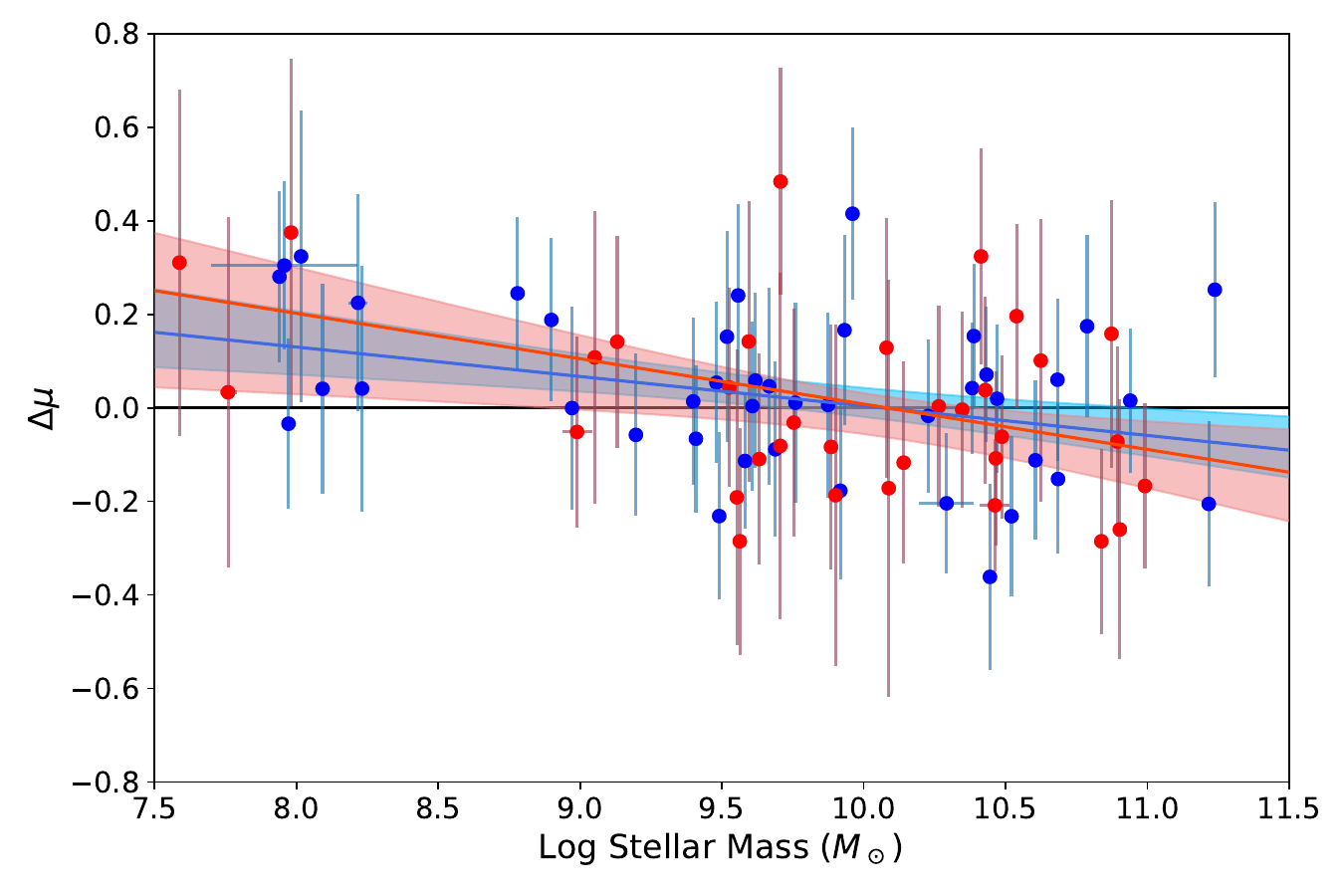}
    \caption{The relationship between Hubble residual and the stellar mass of each SN Ia host galaxy, separated by the colour parameter into red ($c>-0.025$) and blue ($c\leq-0.025$) SNe. The red and blue lines represents the best-fitting relationship for the red and blue samples of SNe Ia respectively, with the shaded regions representing the 1$\sigma$ uncertainty in these fits.}
    \label{fig:mass_vs_mu}
\end{figure}

Many of the stronger correlations with Hubble residual come from the equivalent widths of emission lines. For example, EW[\ion{O}{II}] produces the strongest Pearson correlation coefficient of $r=-0.58 \pm 0.15$, and the slope of the linear fit has a $2.3\sigma$ significance. This is a similar significance to the correlation with host stellar mass with half the sample size, as only half of our observed hosts have a measurable EW [\ion{O}{II}]. The equivalent width of the emission lines we observed, in particular [\ion{O}{II}] and \ion{H}{$\alpha$}, are often used as tracers of the specific star formation rate (sSFR) \citep{bauer_specific_2005, briday_accuracy_2022}.\par

We also investigated a potential relationship between the Hubble residual and the ratio of \ion{H}{$\alpha$} flux to \ion{H}{$\beta$} flux. This ratio of Balmer line fluxes, commonly referred to as the Balmer decrement, has a nominal value of 2.86 for Case B recombination \citep{osterbrock_astrophysics_1989}, and any value greater than this indicates reddening due to dust \citep{dominguez_dust_2013}. The linear fit to this data has a significance of 1.8$\sigma$ (Table \ref{tab:correlation_stats}).

\subsection{Dependence on SN Ia Colour}

\begin{figure}
    \centering
    \includegraphics[width = 0.9\linewidth]{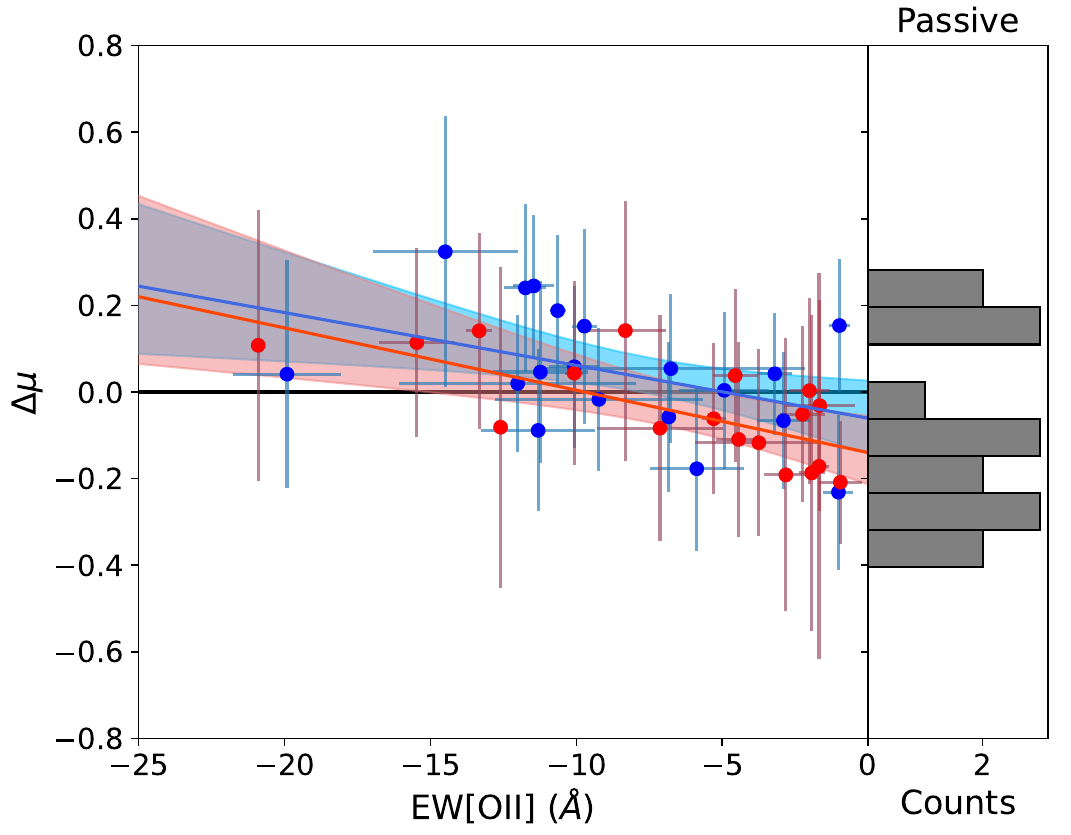}
    \caption{The relationship between Hubble residual and the measured [\ion{O}{II}] equivalent width of each SN Ia host galaxy, separated by the colour parameter into red ($c>-0.025$) and blue ($c\leq-0.025$) SNe. The red and blue lines represents the best-fitting relationship for the red and blue samples of SNe Ia respectively, with the shaded regions representing the 1$\sigma$ uncertainty in these fits. We find no clear dependence of the EW[\ion{O}{II}] relationship on SN colour, with both populations producing slopes of similar magnitude and significance. The histogram on the right shows the distribution in Hubble residual for passive host galaxies, with an EW[\ion{O}{II}] consistent with zero. We find that the passive population largely agrees with the fitted relationship, however we find a small population of passive galaxies with positive Hubble residuals.}
    \label{fig:OII_vs_mu}
\end{figure}

\begin{figure}
    \centering
    \includegraphics[width = 0.9\linewidth]{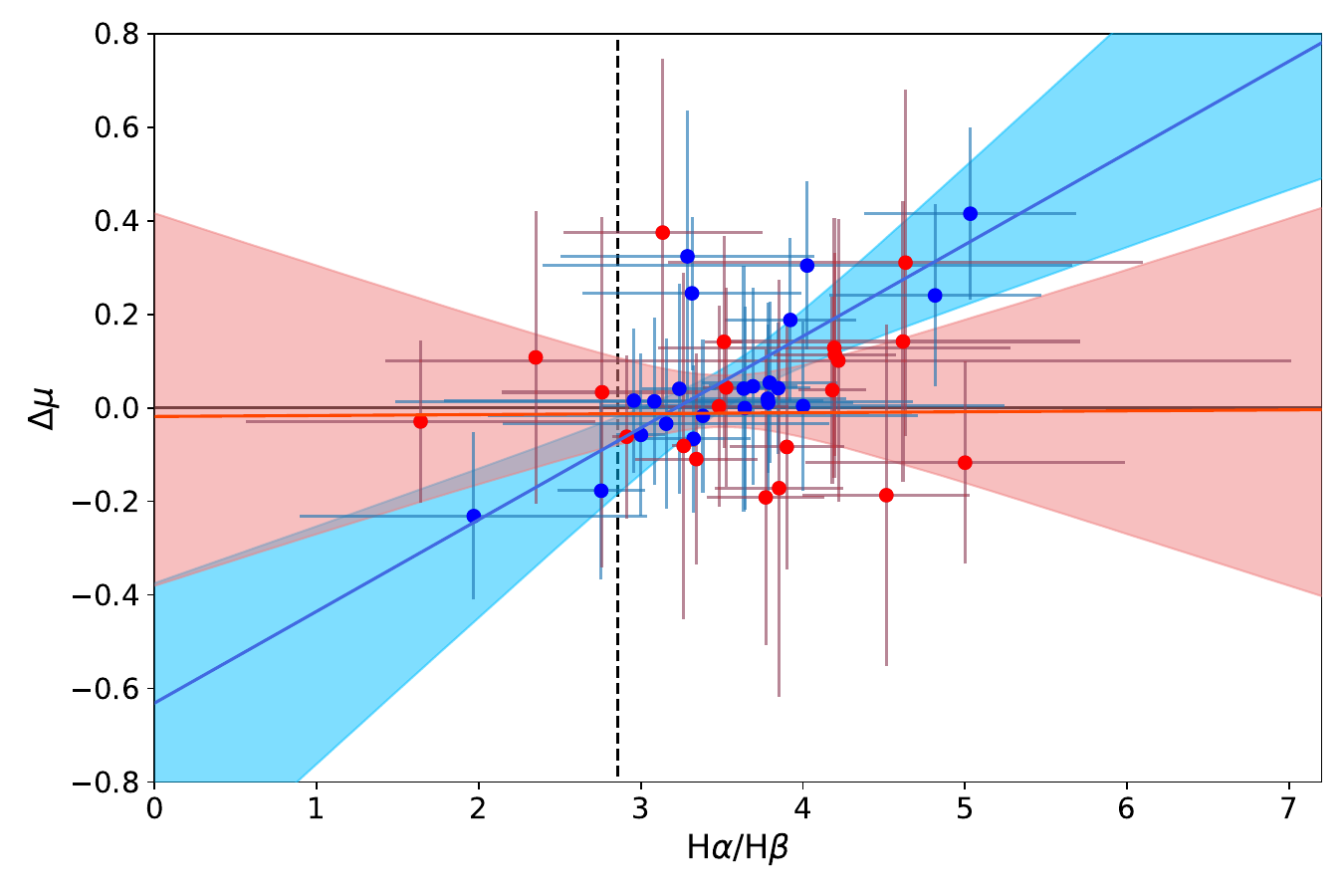}
    \caption{The relationship between Hubble residual and the measured Balmer Decrement of each SN Ia host galaxy, separated by the colour parameter into red ($c>-0.025$) and blue ($c\leq-0.025$) SNe. The red and blue lines represents the best-fitting relationship for the red and blue samples of SNe Ia respectively, with the shaded regions representing the 1$\sigma$ uncertainty in these fits. As in Figure \ref{fig:Included Properties}, the black dashed line represents the theoretical lower limit of the Balmer decrement \citep{osterbrock_astrophysics_1989}. We find that the blue SNe are more tightly correlated with the Balmer decrement, while the red SNe show no correlation.}
    \label{fig:Balmer_vs_mu}
\end{figure}

In order to investigate the interactions at play between the properties of SNe Ia and their host galaxies, we split our host galaxy sample according to the colour of their SNe. This is motivated by the work of \citet{brout_its_2021} (hereafter \citetalias{brout_its_2021}), who find significant differences in the Hubble residual and stellar mass relationship for differently coloured SNe Ia. We utilise the same colour cut as \citetalias{brout_its_2021} and \citet{dixon_using_2022}, where we define red SNe as having a colour parameter $c>-0.025$, while blue supernovae have a colour $c\leq-0.025$.\par

We plot the Hubble residual against host galaxy stellar mass in Figure \ref{fig:mass_vs_mu}, highlighting the blue and red samples of SNe Ia. Table \ref{tab:mass_vs_mu} lists the correlation statistics for these relationships. While we find that the significance of the fitted slope is reduced for each population (owing to a reduced sample size), the slope and correlation coefficients show little variation.\par

\begin{table*}
    \centering
    \caption{The impact of splitting the SN Ia sample into red ($c>-0.025$) and blue ($c\leq-0.025$) SNe on the relationship between host stellar mass and the EW[\ion{O}{II}]. We list the slope and significance of the best-fitting linear relationships, and the Pearson and Spearman correlation coefficients for each subset, along with the results for the combined sample.}
    \begin{tabular}{c c c c c c}
    \hline
    Colour & No. of Hosts & Slope & Significance $(\sigma)$ & Pearson $r$ & Spearman $r$\\
    \hline
    Red & $18$ & $0.053\pm 0.019$ & $2.85$ & $0.61 \pm 0.03$ & $0.53 \pm 0.06$\\
    Blue & $19$ & $0.095\pm 0.030$ & $3.15$ & $0.57 \pm 0.10$ & $0.32 \pm 0.10$\\
    Combined & $37$ & $0.072 \pm 0.016$ & $4.39$ & $0.59 \pm 0.06$ & $0.49 \pm 0.05$\\
    \hline
    \end{tabular}
    
    \label{tab:OII_vs_mass}
\end{table*}

\begin{table*}
    \centering
    \caption{The impact of splitting the SN Ia sample into red ($c>-0.025$) and blue ($c\leq-0.025$) SNe on the relationship between host stellar mass and the Balmer decrement. We list the slope and significance of the best-fitting linear relationships, and the Pearson and Spearman correlation coefficients for each subset, along with the results for the combined sample.}
    \begin{tabular}{c c c c c c}
    \hline
    Colour & No. of Hosts & Slope & Significance $(\sigma)$ & Pearson $r$ & Spearman $r$\\
    \hline
    Red & $21$ & $0.512\pm 0.507$ & $1.01$ & $0.00 \pm 0.21$ & $0.06 \pm 0.17$\\
    Blue & $23$ & $0.410\pm 0.608$ & $0.68$ & $0.06 \pm 0.19$ & $0.11 \pm 0.18$\\
    Combined & $44$ & $0.497\pm 0.430$ & $1.16$ & $0.04 \pm 0.15$ & $0.14 \pm 0.12$\\
    \hline
    \end{tabular}
    
    \label{tab:Balmer_vs_mass}
\end{table*}

\begin{figure}
    \centering
    \includegraphics[width = 0.9\linewidth]{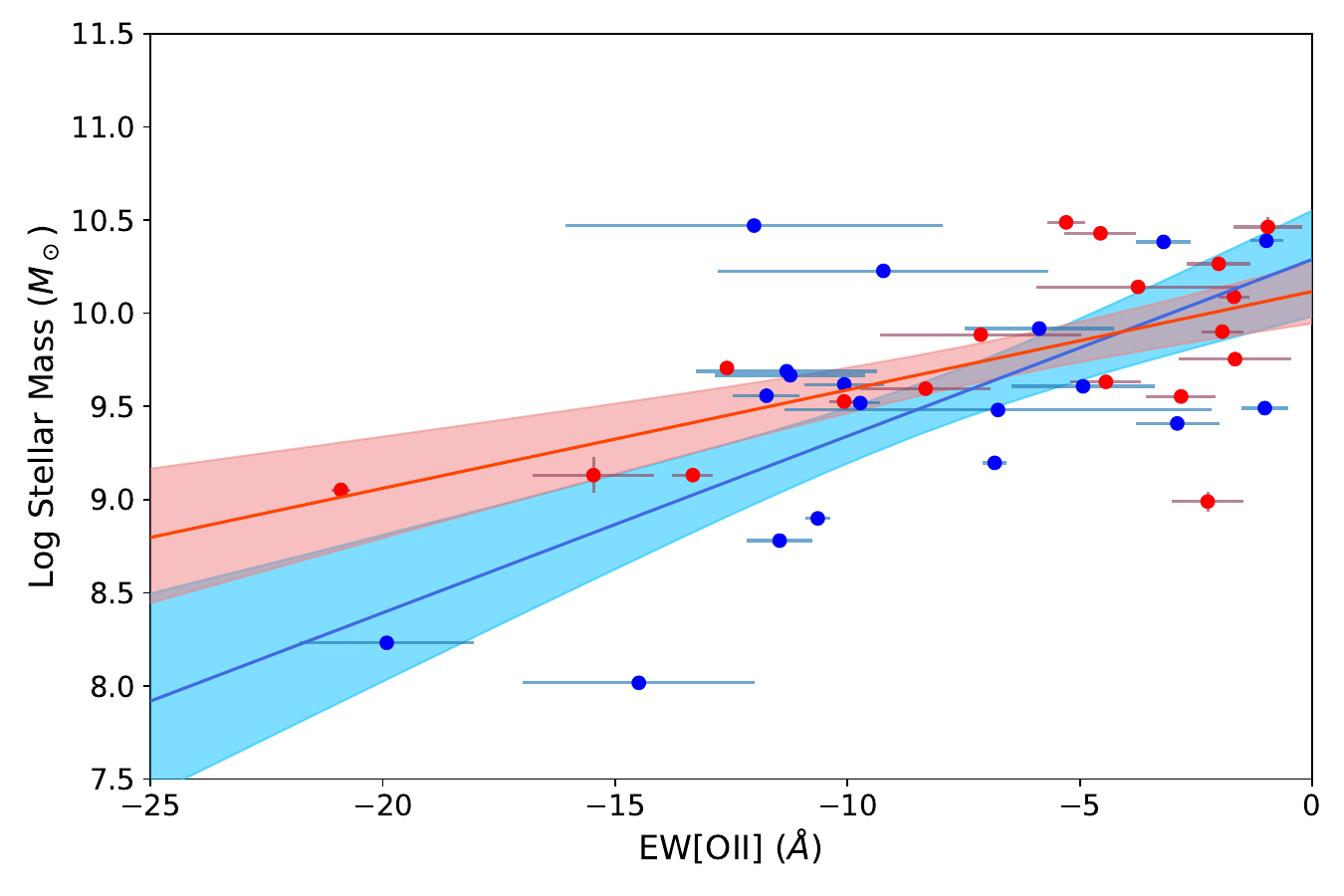}
    \caption{The relationship between stellar mass and the measured [\ion{O}{II}] equivalent width of each SN Ia host galaxy, separated by the colour parameter into red ($c>-0.025$) and blue ($c\leq-0.025$) SNe. The red and blue lines represents the best-fitting relationship for the red and blue samples of SNe Ia respectively, with the shaded regions representing the 1$\sigma$ uncertainty in these fits. We find that both populations appear to show a correlation between these host galaxy properties, with the blue SNe showing a stronger correlation}
    \label{fig:OII_vs_mass}
\end{figure}

\begin{figure}
    \centering
    \includegraphics[width = 0.9\linewidth]{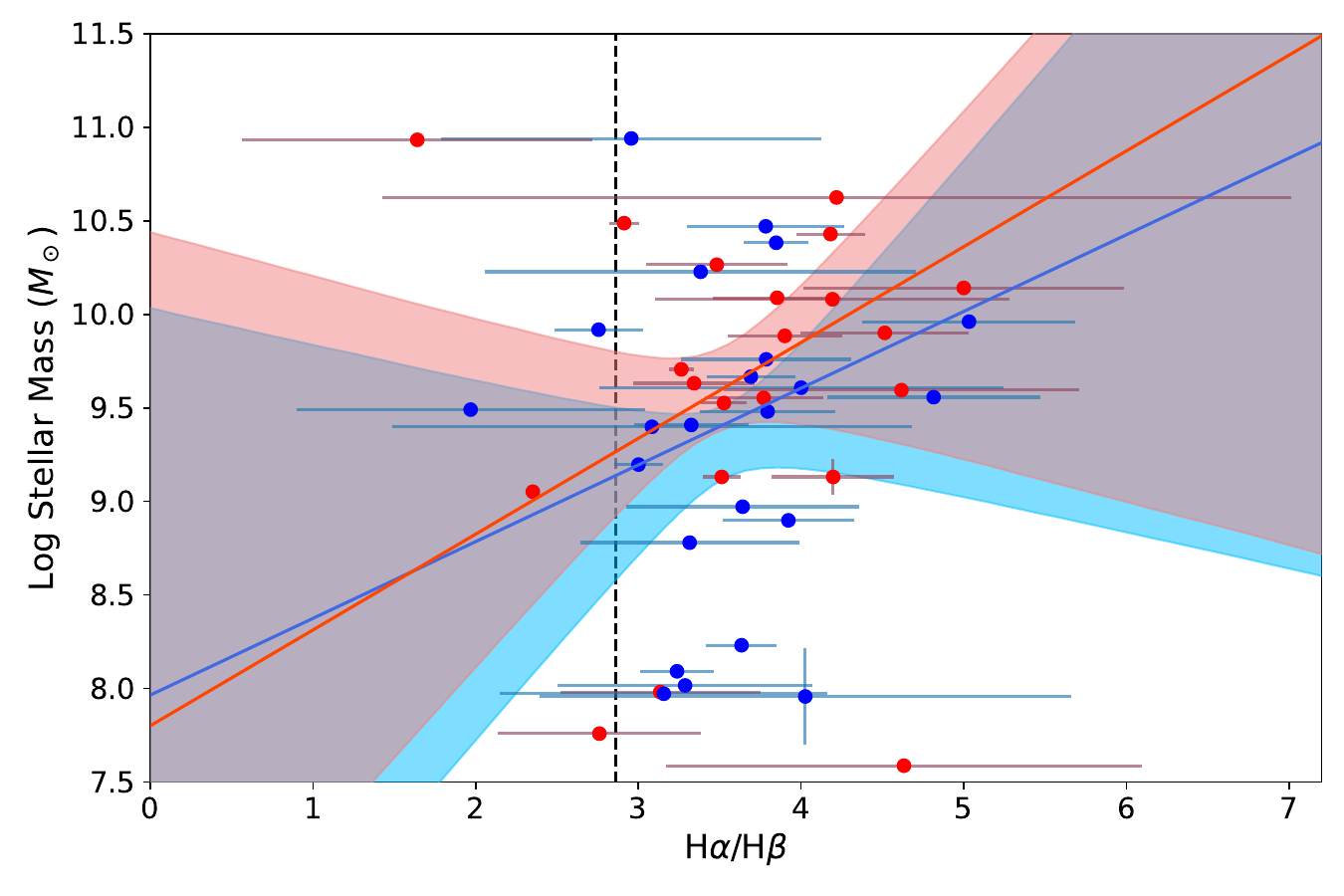}
    \caption{The relationship between stellar mass and the measured Balmer Decrement of each SN Ia host galaxy, separated by the SN colour parameter into red ($c>-0.025$) and blue ($c\leq-0.025$) SNe. The red and blue lines represents the best-fitting relationship for the red and blue samples of SNe Ia respectively, with the shaded regions representing the 1$\sigma$ uncertainty in these fits. Like before, the black dashed line represents the theoretical lower limit of the Balmer decrement \citep{osterbrock_astrophysics_1989}. There is no immediately obvious correlation present.}
    \label{fig:Balmer_vs_mass}
\end{figure}

In Figure \ref{fig:OII_vs_mu} we reproduce the Hubble residual vs. EW[\ion{O}{II}] plot given in Figure \ref{fig:Included Properties}, again juxtaposing the blue and red samples of SNe Ia. Table \ref{tab:OII_vs_mu} lists the correlation statistics for these relationships. We find that the red sample shows a $\sim$1.6$\sigma$ significance in the fitted slope between the Hubble residual and the EW[\ion{O}{II}], while the blue sample's slope has a slightly smaller significance of 1.2$\sigma$. Since \ion{O}{II} equivalent width is a tracer of the specific star formation rate, the plotted points all correspond to SNe Ia from star forming galaxies. We include a histogram on the right showing the distribution in Hubble Residual of SNe Ia coming from passive galaxies, with an EW[\ion{O}{II}] consistent with zero. We find that these SNe generally agree with the fitted relationship for star-forming hosts, however there is a smaller population of SNe from passive galaxies that produce positive Hubble residuals.\par

In Figure \ref{fig:Balmer_vs_mu} we reproduce the Hubble residual vs. Balmer decrement plot given in Figure \ref{fig:Included Properties}, again showing the blue and red samples of SNe Ia. Here we find a mild colour-dependence. We find that the blue SNe Ia Hubble residuals show a 2.2$\sigma$ correlation with the Balmer decrement, while the red SNe are randomly scattered. The correlation statistics for these samples are given in Table \ref{tab:Balmer_vs_mu}. \par

\begin{table}
    \centering
    \rotatebox{90}{
    \begin{minipage}{\textheight}
    \centering
    \caption{The fit parameters and the Pearson and Spearman correlation coefficients for the [\ion{O}{II}] equivalent width and the Hubble residual, for different stages of the light-curve correction. We list the slope and significance of the best-fitting linear relationships, and the Pearson and Spearman correlation coefficients for our sample after applying the $x_1$, $c$ and the Mass (\citetalias{brout_its_2021} $\delta_\mathrm{host}$ and $\delta_\mathrm{bias}$) light-curve corrections. We also include the correlation statistics for the red and blue subsamples for each stage of light-curve correction. The sample contains 37 SNe Ia (19 blue and 18 red) that have a measurable, nonzero EW[\ion{O}{II}] from their host spectrum.}
    \begin{tabular}{c c c c c}
    \hline
    Corrections & Slope & Significance $(\sigma)$ & Pearson $r$ & Spearman $r$\\
    \hline
    $x_1$ & $-0.012\pm 0.008$\textcolor{blue}{($-0.017\pm0.009$)} \textcolor{red}{($-0.014\pm0.012$)}  & 1.44 \textcolor{blue}{(1.84)} \textcolor{red}{(1.23)} & $-0.12 \pm 0.13$ \textcolor{blue}{($-0.52 \pm 0.19$)} \textcolor{red}{($-0.23 \pm 0.20$)} & $-0.16 \pm 0.11$\textcolor{blue}{($-0.54 \pm 0.18$)} \textcolor{red}{($-0.30 \pm 0.19$)}\\
    $x_1 + c$ & $-0.014\pm 0.006$\textcolor{blue}{($-0.012\pm 0.010$)} \textcolor{red}{($-0.014\pm 0.009$)} & 2.30 \textcolor{blue}{(1.24)} \textcolor{red}{(1.61)} & $-0.58 \pm 0.15$ \textcolor{blue}{($-0.45 \pm 0.20$)} \textcolor{red}{($-0.70 \pm 0.23$)} & $-0.58 \pm 0.14$\textcolor{blue}{($-0.43 \pm 0.20$)} \textcolor{red}{($-0.68 \pm 0.21$)}\\
    $x_1 + c + \mathrm{Mass}$ & $-0.012\pm 0.006$\textcolor{blue}{($-0.013\pm0.008$)} \textcolor{red}{($-0.011\pm0.009$)} & 1.98 \textcolor{blue}{(1.54)} \textcolor{red}{(1.24)} & $-0.41 \pm 0.16$ \textcolor{blue}{($-0.45 \pm 0.20$)} \textcolor{red}{($-0.35 \pm 0.23$)} & $-0.42 \pm 0.15$\textcolor{blue}{($-0.39 \pm 0.20$)} \textcolor{red}{($-0.41 \pm 0.23$)}\\
    \hline
    \end{tabular}
    \end{minipage}}
    \label{tab:OII_stages}
\end{table}

\begin{table}
    \centering
    \rotatebox{90}{
    \begin{minipage}{\textheight}
    \centering
    \caption{The fit parameters and the Pearson and Spearman correlation coefficients for the Balmer decrement and the Hubble residual, for different stages of the light-curve correction. We list the slope and significance of the best-fitting linear relationships, and the Pearson and Spearman correlation coefficients for our sample after applying the $x_1$, $c$ and the Mass(\citetalias{brout_its_2021} $\delta_\mathrm{host}$ and $\delta_\mathrm{bias}$) light-curve corrections. We also include the correlation statistics for the red and blue subsamples for each stage of light-curve correction. The sample contains 44 SNe Ia (23 blue and 21 red) that have a measurable Balmer decrement from their host spectrum.}
    \begin{tabular}{c c c c c}
    \hline
    Corrections & Slope  & Significance $(\sigma)$ & Pearson $r$ & Spearman $r$\\
    \hline
    $x_1$ & $0.171\pm 0.144$\textcolor{blue}{($0.168\pm0.096$)} \textcolor{red}{($0.085\pm0.172$)} & 1.19 \textcolor{blue}{(1.75)} \textcolor{red}{(0.50)} & $0.35 \pm 0.15$ \textcolor{blue}{($0.57 \pm 0.19$)} \textcolor{red}{($0.34 \pm 0.21$)} & $0.36 \pm 0.13$\textcolor{blue}{($0.41 \pm 0.19$)} \textcolor{red}{($0.40 \pm 0.20$)}\\
    $x_1 + c$ & $0.135\pm0.077$\textcolor{blue}{($0.196\pm0.090$)} \textcolor{red}{($0.002\pm0.117$)} & 1.75 \textcolor{blue}{(2.19)} \textcolor{red}{(0.02)} & $0.33 \pm 0.14$ \textcolor{blue}{($0.72 \pm 0.19$)} \textcolor{red}{($0.01 \pm 0.21$)} & $0.31 \pm 0.14$\textcolor{blue}{($0.61 \pm 0.18$)} \textcolor{red}{($0.05 \pm 0.21$)}\\
    $x_1 + c + \mathrm{Mass}$ & $0.111\pm0.088$\textcolor{blue}{($0.191\pm0.085$)} \textcolor{red}{($0.015\pm0.127$)} & 1.27 \textcolor{blue}{(2.25)} \textcolor{red}{(0.12)} & $0.41 \pm 0.14$ \textcolor{blue}{($0.72 \pm 0.18$)} \textcolor{red}{($0.16 \pm 0.22$)} & $0.39 \pm 0.14$\textcolor{blue}{($0.57 \pm 0.18$)} \textcolor{red}{($0.23 \pm 0.21$)}\\
    \hline
    \end{tabular}
    \end{minipage}}
    \label{tab:Balmer_stages}
\end{table}

\subsection{Dependence on Host Stellar Mass}

Galaxy stellar mass correlates with many other galaxy observables. For example, mass correlates with specific star formation rate \citep{bauer_specific_2005} and dust properties \citep{dominguez_dust_2013,salim_dust_2018}.\par

In Figure \ref{fig:OII_vs_mass}, we plot the EW[\ion{O}{II}] of our host galaxies against their stellar mass, again separating by SNe Ia colour. We find a strong 4.4$\sigma$ correlation between these properties, which agrees with the idea that low-mass galaxies have larger sSFRs, and therefore stronger [\ion{O}{II}] equivalent widths. The statistics of these correlations are given in Table \ref{tab:OII_vs_mass}, and we find that while both populations produce a correlation, the blue sample's slope is steeper and more significant.\par

We plot the Balmer decrement of our host galaxies against their stellar mass in Figure \ref{fig:Balmer_vs_mass}, again separating by SNe Ia colour. We expect that we should find a dependence between these quantities since more massive galaxies have been shown to have greater amounts of dust, but shallower attenuation laws \citep{salim_dust_2018}. We find a 1.2$\sigma$ linear fit between these properties, which is not significant enough to claim the existence of a correlation. The statistics of these linear fits are given in Table \ref{tab:Balmer_vs_mass}. \par

\section{Discussion} \label{sec:Discussion}

\begin{figure*}
    \centering
    \includegraphics[width = \linewidth]{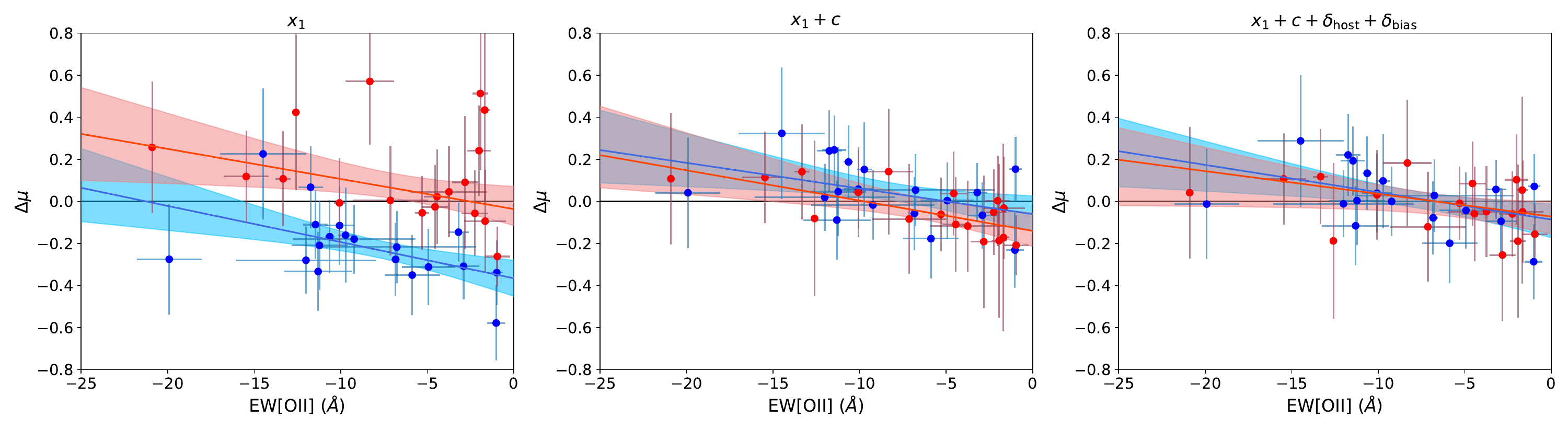}
    \caption{Left to Right: Plots of the Hubble residual against the [\ion{O}{II}] equivalent width after implementing the light-curve width ($x_1$), colour ($c$) and the \citetalias{brout_its_2021} host-mass and bias corrections ($\delta_\mathrm{host}$ and $\delta_\mathrm{bias}$). The data points are coloured according to the classification of their SNe Ia as blue ($c\leq-0.025$) or red ($c>-0.025$). Note how the correlation between Hubble residual and EW of [\ion{O}{II}] persists even after the mass-step and bias corrections are applied.}
    \label{fig:OII_stages}
\end{figure*}

\begin{figure*}
    \centering
    \includegraphics[width = \linewidth]{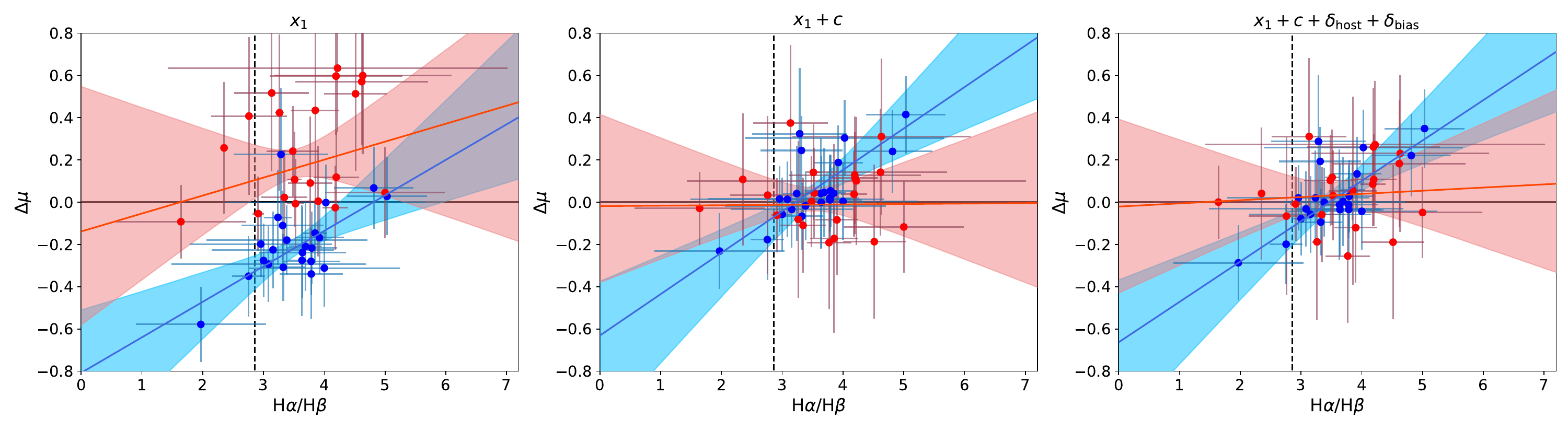}
    \caption{Left to Right: Plots of the Hubble residual against the Balmer decrement after implementing the light-curve width ($x_1$), colour ($c$) and the \citetalias{brout_its_2021} host-mass and bias corrections ($\delta_\mathrm{host}$ and $\delta_\mathrm{bias}$). The black dashed line indicates the theoretical lower limit of the Balmer decrement \citep{osterbrock_astrophysics_1989}. The data points are coloured according to the classification of their SNe Ia as blue ($c\leq-0.025$) or red ($c>-0.025$). Note how the correlation between Hubble residual and Balmer decrement persists even after the mass-step and bias corrections are applied, but only for the blue SNe Ia.}
    \label{fig:Balmer_stages}
\end{figure*}

\subsection{The Balmer decrement as a measure of dust attenuation}  

We have shown that the Hubble residual correlates with host galaxy mass and with the sSFR, and that the sSFR correlates with host galaxy mass. But, what is the physical cause of the correlations with the Hubble residual. Might it be the simplistic treatment of the SN Ia colour correction in the \citet{tripp_two-parameter_1998} equation (Equation \ref{eq:Tripp}), and its inability to capture the effects of dust attenuation, as posited by \citetalias{brout_its_2021}?\par 

In their recent work, \citetalias{brout_its_2021} propose a new model, by breaking the observed colour into two components: an intrinsic colour $c_\mathrm{int}$ associated with SN Ia properties, and a dust component $E_\mathrm{dust}$. The change in observed brightness is thus modelled as:

\begin{equation}
    \Delta m_B = \beta_\mathrm{SN} c_\mathrm{int} + (R_V+1)E_\mathrm{dust} + \epsilon_\mathrm{noise}
\end{equation}\par

\noindent where $\beta_\mathrm{SN}$ is derived from the intrinsic colour-luminosity relationship of SNe Ia, $R_V$ is the total-to-selective extinction ratio and $\epsilon_\mathrm{noise}$ is the measurement error. Using this model, \citetalias{brout_its_2021} add a dependence between dust properties and host stellar mass. They split their $R_V$ and $E_\mathrm{dust}$ parameters into a high-mass ($>10^{10} M_\odot$) and low-mass ($<10^{10} M_\odot$) regime, and find that this better describes the data. The $\beta_\mathrm{SN}$ and $c_\mathrm{int}$ parameters however are not split, as they relate to intrinsic properties of SNe Ia and have no dependence on host environment.\par

\citetalias{brout_its_2021} demonstrate the improved capability of their model by testing it against a combination of high and low-redshift data from the Carnegie Supernova Project \citep[CSP;][]{stritzinger_distance_2010}, CfA3-4 \citep{hicken_cfa3_2009, hicken_cfa4_2012}, PAN-STARRS1 \citep{rest_cosmological_2014, scolnic_complete_2018}, the Sloan Digital Sky Survey \citep{sako_photometric_2011}, the Supernova Legacy Survey \citep{betoule_improved_2014}, the Dark Energy Survey 3-year sample \citep{brout_first_2019}, and the Foundation sample \citep{foley_foundation_2018, jones_foundation_2019}. Using their model, the authors are able to reproduce the RMS scatter and mean Hubble residual as a function of SN Ia colour. These relationships are dependent on host stellar mass, and the \citetalias{brout_its_2021} model is consistent with observations in both the high and low mass regimes. The authors also find that the host galaxy mass-step is a function of colour, and can be reproduced by their updated dust model. In summary, the overly simplistic approach used to correct for the correlation between colour and luminosity adopted by current light-curve fitters such as SALT3 \citep{Kenworthy_salt3_2021} may be the driving mechanism behind observed correlations with the Hubble residual, including the host galaxy mass-step. \par

We can investigate the impact of dust in our own dataset by determining the Balmer decrement. The Balmer decrement, defined as the ratio of \ion{H}{$\alpha$} flux to \ion{H}{$\beta$} flux, is a useful indicator of the amount of reddening due to dust in the host galaxy \citep{osterbrock_astrophysics_1989, dominguez_dust_2013}. In Figure \ref{fig:Included Properties} we show the relationship between the Hubble residual and the Balmer decrement. The linear fit to this relationship has a $\sim$1.8$\sigma$ significance (Table \ref{tab:correlation_stats}), which is too low to claim a correlation. Further data will be needed to confirm this trend.\par

By comparing galaxies that host red SNe with those that host blue SNe, we reveal a tentative difference -- the blue sample's Hubble residuals are correlated with the Balmer decrement, while the red sample is uncorrelated (Figure \ref{fig:Balmer_vs_mu}). We expect that dustier galaxies will produce intrinsically fainter SNe. Our results suggest that the red SNe are accurately corrected for dust attenuation, while the blue SNe are slightly under-corrected.\par

When comparing the Balmer decrement with host stellar mass, we expect to find a positive correlation, i.e. high mass star-forming galaxies are more attenuated by dust \citep{salim_dust_2018}. In Figure \ref{fig:Balmer_vs_mass} we find that while there is a slight trend, the trend is not significant. More data is required to confirm this correlation. \par

In the \citetalias{brout_its_2021} model, the authors find that more massive galaxies have lower values of $R_V$, and therefore steeper attenuation laws. However, \citet{salim_dust_2018} show that dust properties depend on both host galaxy mass and sSFR. At constant mass, passive galaxies have a shallower attenuation law than star-forming galaxies on the main sequence, as do low mass star forming galaxies. \citetalias{brout_its_2021} do not split their high mass sample into galaxies that are passive or star-forming, therefore it is not straightforward to compare \citetalias{brout_its_2021} with \citet{salim_dust_2018}. \par

By comparing the EW[\ion{O}{II}] correlations in Figures \ref{fig:OII_vs_mu} and \ref{fig:OII_vs_mass}, we can determine whether our results are consistent with the \citetalias{brout_its_2021} model. When looking at the red SNe Ia only (Fig. 6 of \citetalias{brout_its_2021}), SNe Ia from low-mass galaxies have positive Hubble residuals, while the SNe Ia from high mass galaxies have negative residuals. In Figure \ref{fig:OII_vs_mass}, low-mass galaxies with red SNe have large [\ion{O}{II}] equivalent widths, and in Figure \ref{fig:OII_vs_mu}, large equivalent widths give positive residuals. Following the same logic, high mass galaxies in Figure \ref{fig:OII_vs_mass} have small [\ion{O}{II}] equivalent widths, which in turn give negative Hubble residuals in Figure \ref{fig:OII_vs_mu}. This is consistent with the behaviour seen in \citetalias{brout_its_2021}. However, we find that blue SNe show the same relationship in Figure \ref{fig:OII_vs_mass}, with high mass galaxies having small [\ion{O}{II}] equivalent widths and therefore negative Hubble residuals. In Fig. 6 of \citetalias{brout_its_2021}, the authors find that blue SNe have, on average, positive hubble residuals with little dependence on mass. Therefore our findings with EW [\ion{O}{II}] measurements are not entirely consistent with the findings of \citetalias{brout_its_2021}. This inconsistency may be caused by the secondary dependence of dust attenuation on sSFR, traced by the EW [\ion{O}{II}], meaning that this dependence will need to be considered in future intrinsic scatter models in order to more accurately correct individual SN Ia light-curves. However, the trends in the data are weak. More data will be needed to confirm the trends seen here.

\begin{figure*}
    \centering
    \includegraphics[width = 0.8\linewidth]{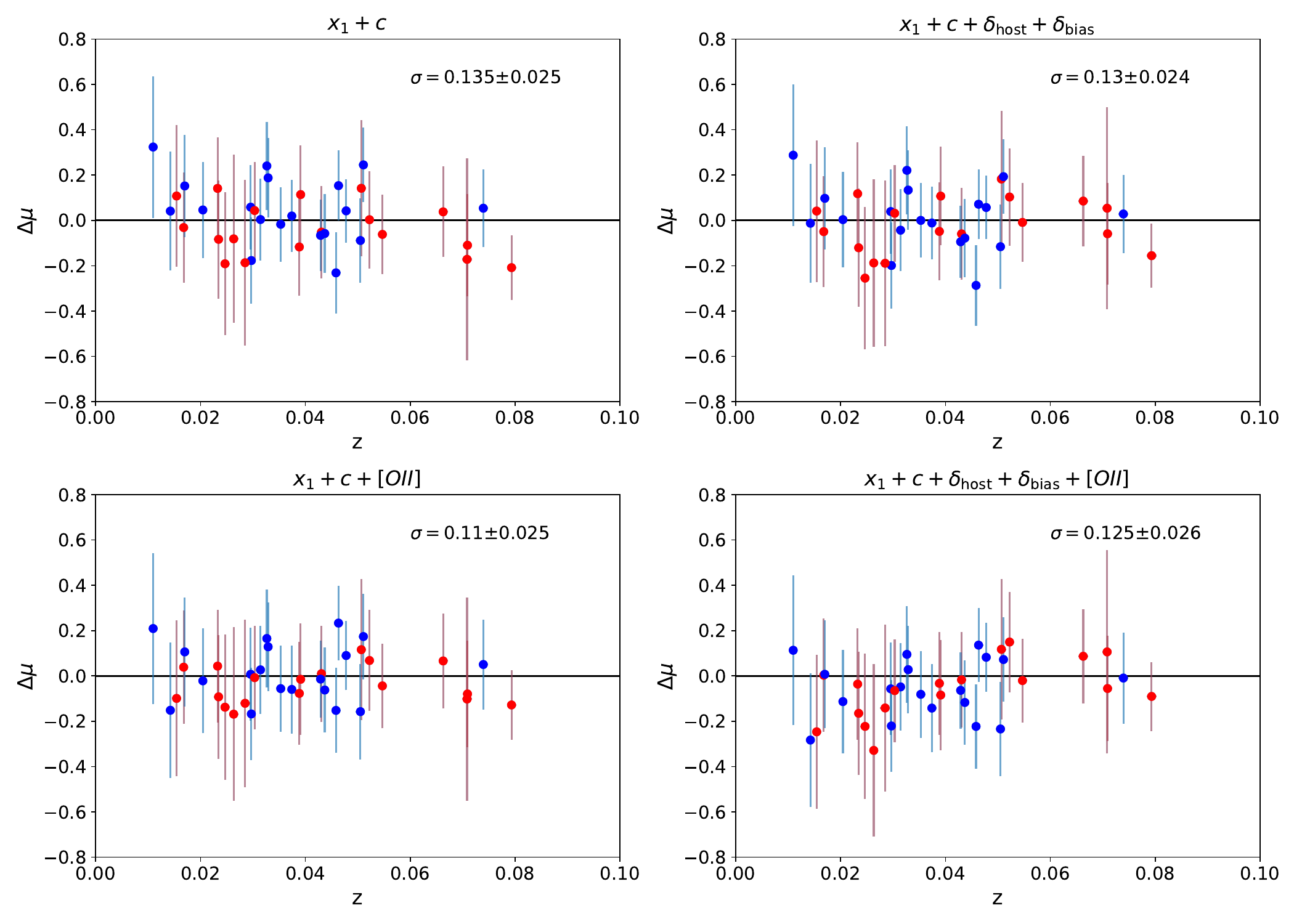}
    \caption{Plots of the Hubble residual against redshift for four different cases of SN Ia light-curve correction. Along the top row we show the canonical corrections for stretch ($x_1$), colour ($c$), and the \citetalias{brout_its_2021} host-mass and bias corrections ($\delta_\mathrm{host}$ and $\delta_\mathrm{bias}$). The bottom row shows the same plots, but with an added correction for the [\ion{O}{II}] equivalent width. We only plot the points of SNe with a measurable EW [\ion{O}{II}] in their host spectra, and the data points are coloured according to the classification of the SNe Ia as blue ($c\leq-0.025$) or red ($c>-0.025$). The resulting scatter in Hubble residual ($\sigma$) is shown for each of the four cases. We find that while the mass and bias corrections reduce the scatter in the Hubble diagram, the EW [\ion{O}{II}] correction alone reduces the scatter more.}
    \label{fig:scatters}
\end{figure*}

\subsection{Impact of the Mass Step Correction} \label{sec:Applying Mass Step}

We posit that the incomplete treatment of dust in SN Ia colour corrections is the driving factor behind the correlations that we see, including the mass-step. By correcting for these correlations, we can lessen the impact of systematic errors on SN Ia distance measurements, and reduce the scatter in the Hubble diagram.\par

When accounting for the host galaxy mass-step, the \citet{tripp_two-parameter_1998} equation (Equation \ref{eq:Tripp}) is often modified to have the following form:
\begin{equation}\label{eq:Tripp_with_Mass}
    \mu_\mathrm{obs} = m_B - M_B + \alpha x_1 - \beta c + \delta_\mathrm{host} - \delta_\mathrm{bias}
\end{equation}

\noindent where the added $\delta_\mathrm{bias}$ term corrects for selection biases in the sample, and $\delta_\mathrm{host}$ provides the mass-step correction \citep{brout_pantheon_2022},
\begin{equation}\label{eq:Mass_Step_Term}
    \delta_\mathrm{host} = \gamma \times \left(1 + e^{(M_* - S)/\tau_{M_*}}\right)^{-1} - \frac{\gamma}{2}
\end{equation}

\noindent where $\gamma$ represents the size of the mass-step (in magnitudes), $M_*$ is the stellar mass of the host galaxy in solar units ($M_\odot$), $S$ is the location of the mass step (canonically $S=10^{10}M_\odot$), and $\tau_{M_*}$ is the width of the mass step.\par

If the correlations we see with host properties, including the mass step, are driven by the treatment of dust as outlined by \citetalias{brout_its_2021}, then upon applying their host-mass and bias corrections, our observed correlations should disappear.\par

We show plots of the Hubble residual against the [\ion{O}{II}] equivalent width for different stages of light-curve correction in Figure \ref{fig:OII_stages}, with the linear fit parameters and Pearson and Spearman correlation coefficients for the red, blue and combined populations listed in Table \ref{tab:OII_stages}. We find that for the blue population of SNe, the $1.8\sigma$ dependence on EW[\ion{O}{II}] weakens after the colour correction is applied. When looking at the red population, the significance of the fitted slope strengthens upon applying the colour correction. The overall correlation for the combined population does not reduce significantly after the \citetalias{brout_its_2021} mass-step and bias corrections are applied, suggesting that these corrections are not fully accounting for the overly simplistic colour correction in the \citet{tripp_two-parameter_1998} formula (Equation \ref{eq:Tripp}).\par

We show plots of the Hubble residual against Balmer decrement for different stages of light-curve correction in Figure \ref{fig:Balmer_stages}, with the linear fit parameters and Pearson and Spearman correlation coefficients for the red, blue and combined populations listed in Table \ref{tab:Balmer_stages}. We find that for the blue population of SNe, the correlation again remains after all three stages of correction (light-curve width $x_1$, colour $c$ and the mass-step and bias corrections $\delta_\mathrm{host}$ and $\delta_\mathrm{bias}$), while for the red population, no relationships are found at any stage.\par

When we correct for the host galaxy stellar mass dependence following \citet{brout_pantheon_2022}, we find that this does not entirely remove the trend in the Hubble residuals with either EW [\ion{O}{II}] or the Balmer decrement. This suggests that we need to look beyond the mass-step if we wish to improve SN Ia distance measurements.\par

\subsection{Applying an [\ion{O}{II}] Correction}

While corrections for the host galaxy mass step have been proven effective at reducing the scatter in the Hubble diagram, we still see correlations between the Hubble residual and other host properties even after the mass step and bias corrections have been applied. In Table \ref{tab:correlation_stats} we find that the linear fit with the [\ion{O}{II}] equivalent width is nearly as strong as the mass correlation, with equivalent intrinsic scatters about these fits. An EW[\ion{O}{II}] correction would be particularly favourable, as [\ion{O}{II}] is a bright spectral feature that is observable in optical/NIR spectra out to $z\sim 1.5$. As host spectra are already obtained for the purposes of determining supernova redshifts, the EW[\ion{O}{II}] could be extracted from these spectra to provide the correction to a SN Ia distance modulus. We have therefore considered the impact of applying a correction for the EW [\ion{O}{II}] correlation, and compared the impact of this correction to the established mass step correction that has been applied for the last decade.\par

In Figure \ref{fig:scatters} we plot Hubble residuals at different stages of correction against redshift, displaying the resulting scatter $\sigma$ in the Hubble diagram after these corrections have been applied. In the top panels, we apply the standard stretch and colour corrections, followed by the inclusion of the canonical host mass and bias corrections set by the \citetalias{brout_its_2021} model. In the bottom panels we apply a correction for our observed EW [\ion{O}{II}] correlations. On the left we apply the correction according to the combined slope of the EW [\ion{O}{II}] relationship ($m=-0.014\pm0.006$) in  Table \ref{tab:correlation_stats}, while in the right panel we apply the EW [\ion{O}{II}] correction according to the remaining correlation after the SN Ia sample has been corrected for mass and selection biases ($m = -0.012 \pm 0.006$, Table \ref{tab:OII_stages}).\par

We find that upon applying the EW [\ion{O}{II}] correction alone, we are able to reduce the scatter in the Hubble diagram to a greater extent than is achieved by the mass-step and bias corrections. When the EW [\ion{O}{II}] and the \citetalias{brout_its_2021} mass and bias corrections are all applied, the scatter is still reduced, but to a lesser extent. We note that in all four panels of Figure \ref{fig:scatters}, only those SNe with a measurable EW [\ion{O}{II}] from their host spectra are included. \par

\begin{figure*}
    \centering
    \includegraphics[width = 0.8\linewidth]{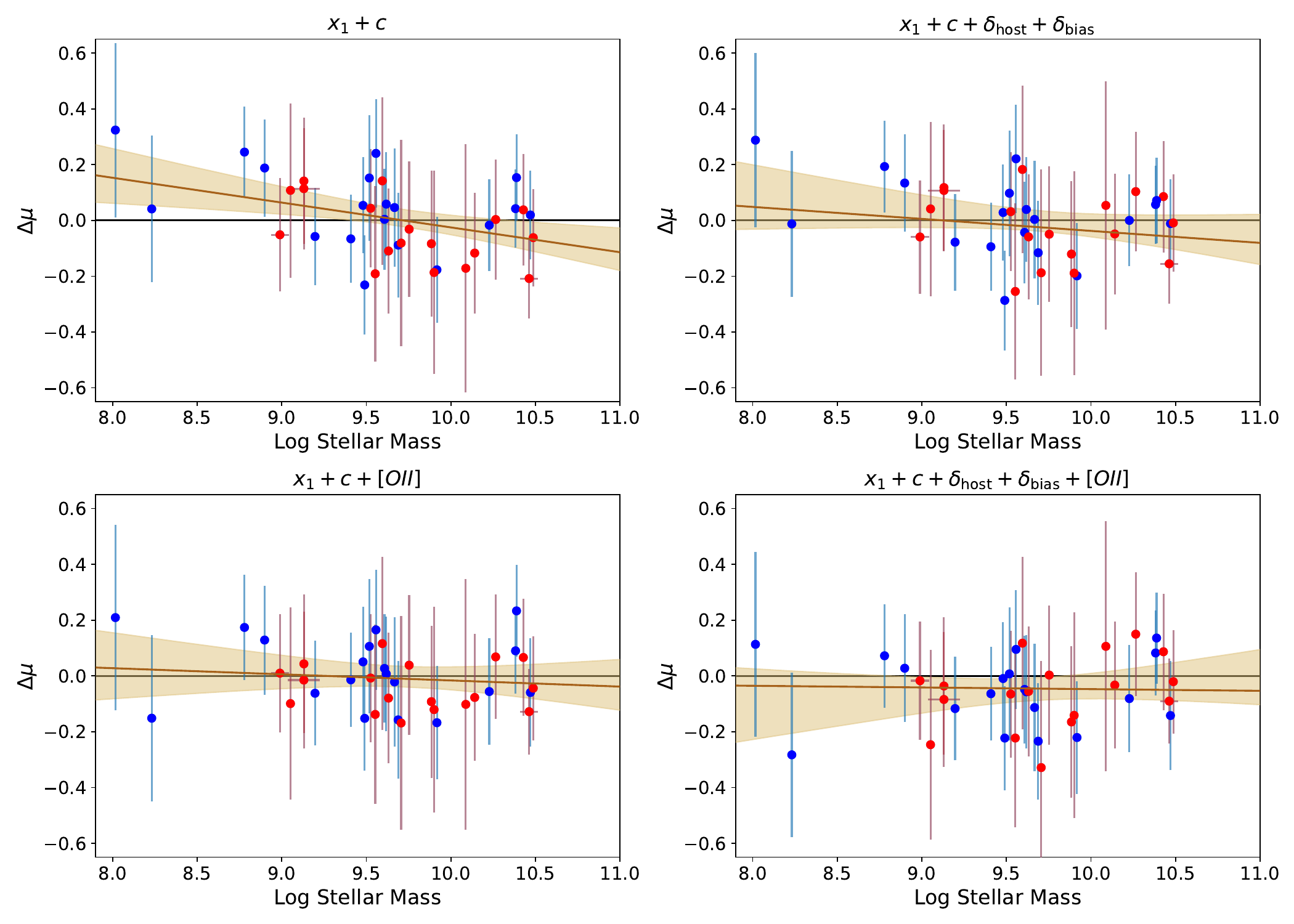}
    \caption{Plots of the Hubble residual against the host stellar mass after implementing the light-curve width ($x_1$) and colour ($c$), mass-step and bias ($\delta_\mathrm{host} + \delta_\mathrm{bias}$) and EW [\ion{O}{II}] corrections to the Hubble residual. The data points are coloured according to the classification of their SNe Ia as blue ($c\leq-0.025$) or red ($c>-0.025$). The gold line represents the best-fitting linear relationship to the combined dataset, with the shaded region representing a 1$\sigma$ confidence interval. Only those SNe with a measurable EW [\ion{O}{II}] from their host spectrum are shown. We find that the EW [\ion{O}{II}] correction is able to remove the host mass relationship before a mass correction has even been applied.}
    \label{fig:mass_OII_corr}
\end{figure*}

\begin{figure*}
    \centering
    \includegraphics[width = 0.8\linewidth]{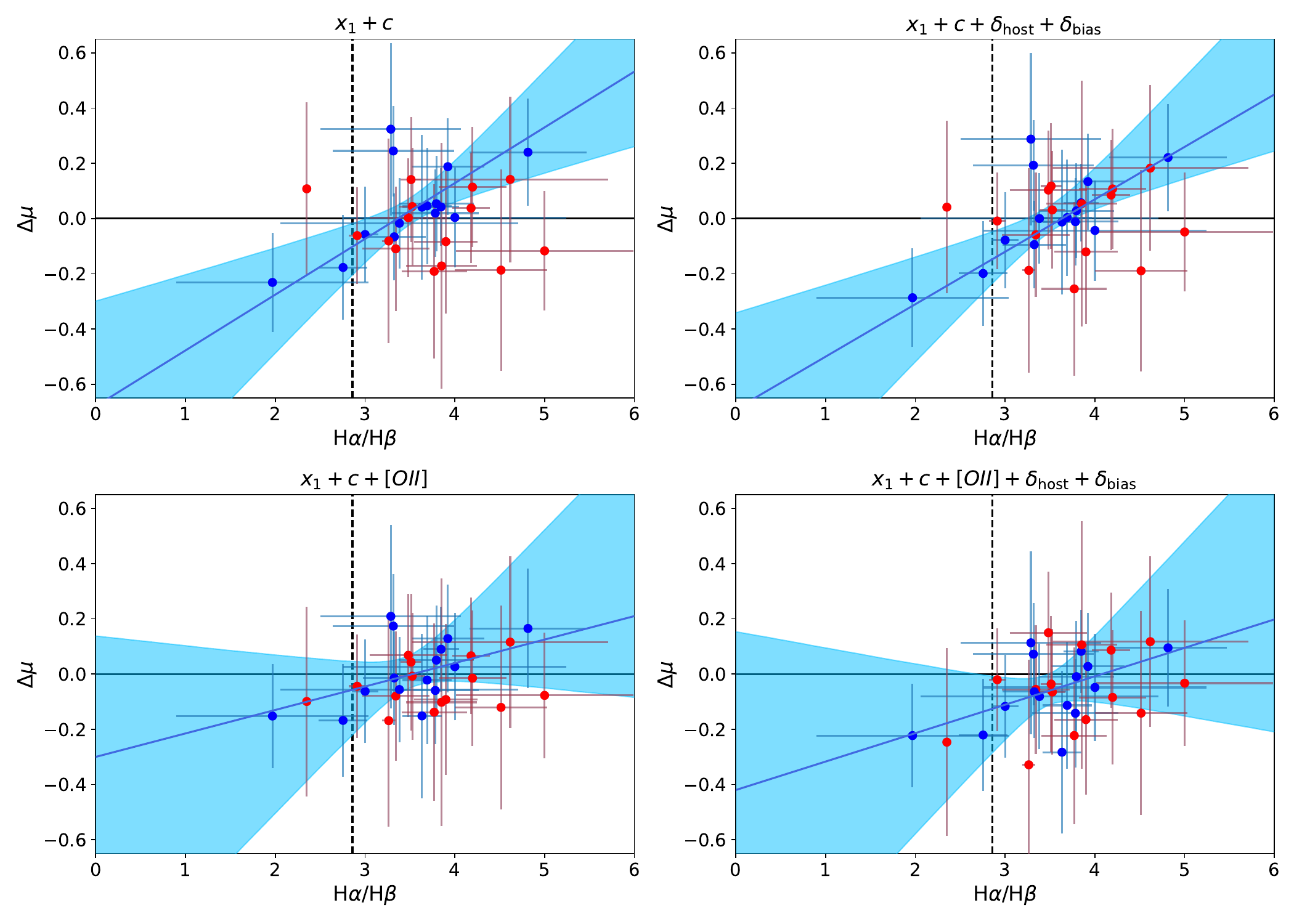}
    \caption{Plots of the Hubble residual against the Balmer decrement after implementing the light-curve width ($x_1$) and colour ($c$), mass-step and bias ($\delta_\mathrm{host} + \delta_\mathrm{bias}$) and EW [\ion{O}{II}] corrections to the Hubble residual. The black dashed line indicates the theoretical lower limit of the Balmer decrement \citep{osterbrock_astrophysics_1989}. The data points are coloured according to the classification of their SNe Ia as blue ($c\leq-0.025$) or red ($c>-0.025$). The blue line represents the best-fitting linear relationship to the blue sample of SNe Ia, with the shaded region representing a 1$\sigma$ confidence interval. Only those SNe with a measurable EW [\ion{O}{II}] from their host spectrum are shown. We find that the EW [\ion{O}{II}] correction is able to greatly reduce the significance of the linear fit} between the Balmer decrement and the Hubble residual of blue SNe Ia, which is not achieved by the host mass correction.
    \label{fig:Balmer_OII_corr}
\end{figure*}

In Figure \ref{fig:mass_OII_corr}, we plot the Hubble residuals at different stages of correction against host stellar mass, using the same corrections as Figure \ref{fig:scatters}. The resulting correlation statistics are shown in Table \ref{tab:mass_OII_corr}. We find that by applying an EW [\ion{O}{II}] correction, we are able to greatly reduce the significance of the host galaxy mass relationship before bias and mass-step corrections have been applied. As we saw in Figure \ref{fig:OII_stages}, applying the mass step correction did not in turn remove our EW [\ion{O}{II}] correlation, suggesting that the EW [\ion{O}{II}] is more closely linked to the overarching systematic driver of the scatter in the Hubble diagram.\par

We also investigated the impact of the EW [\ion{O}{II}] correction on our observed correlation between the Hubble residuals of blue SNe Ia and their Balmer decrements, which again remains after a mass step correction is applied (see Figure \ref{fig:Balmer_stages}). In Figure \ref{fig:Balmer_OII_corr} we apply light-curve corrections in the same way as Figures \ref{fig:scatters} and \ref{fig:mass_OII_corr}, and find that the EW [\ion{O}{II}] correction is able to break down the relationship between the Hubble residual of blue SNe Ia and the Balmer decrement, reducing the slope and the significance of the linear fit from $\sim 1.7\sigma$ to $\sim 0.2\sigma$ (see Table \ref{tab:Balmer_OII_corr}). This provides further support to the idea that [\ion{O}{II}] is more closely related to the central driver of scatter in the Hubble diagram, and that correlations with host stellar mass and the Balmer decrement are by-products of this overarching systematic influence.\par

We find that an EW[\ion{O}{II}] correction to SN Ia distance moduli has an equivalent, and potentially greater impact on reducing the scatter in the Hubble diagram than the mass-step and bias corrections, and unlike these corrections, is able to account for correlations seen with other host galaxy properties. We therefore suggest that correcting Hubble residuals for [\ion{O}{II}] equivalent width in place of or in addition to host stellar mass may prove to be a viable means of reducing the scatter in the Hubble diagram. This may more accurately account for the limitations of the single SN Ia colour parameter in its ability to account for the impact of dust attenuation along the line of sight.\par

\subsection{Future Work and Applications}

While we have identified a number of correlations between the Hubble residual and SN Ia host galaxy properties, there are limitations of our work.\par

The main limitation of our work is the small sample size when assessing the colour dependence of our results. Many of the linear fits we have applied to our data have a significance of $1-2\sigma$, particularly when analysing relationships with the Balmer decrement. We therefore need a much larger sample of SNe Ia and their host galaxies in order to make conclusions about these relationships. There remain $\sim 100$ SNe from Foundation \citep{foley_foundation_2018, jones_foundation_2019}, along with a range of other low-redshift SNe Ia datasets whose host galaxies could be observed and analysed.\par

Another limitation of our work is that we have derived global properties of our SN Ia host galaxies. A strong step-like dependence of the Hubble residual has been shown when analysing the local-sSFR \citep{rigault_strong_2020,briday_accuracy_2022}. By utilising IFU spectroscopy, spectra extracted from a smaller aperture centreed on the location of each SN might provide a clearer result than using spectra of the entire host \citep[e.g.][]{rigault_evidence_2013,kelsey_effect_2021}.\par   

While there are limitations of our research, there are many ways in which this work could be built upon and applied in practice when constraining cosmological parameters with SN Ia distance measurements. Firstly, we can consider other functional forms of the host galaxy stellar mass relationship with Hubble residual. Treating this relationship as a step function may be the reason why the mass-step correction does not fully account for our observed correlations between Hubble residual and host properties. \citet{dixon_using_2022} apply a linear correction for stellar mass, rather than collapsing a step function, and we fit a linear relationship with stellar mass with a confidence of $\sim$1.6$\sigma$ in our own analysis (Table \ref{tab:correlation_stats}), however other functional forms could also be considered. 

Another area of future investigation is to investigate the gas-phase metallicity of our host galaxies. Ratios of the emission lines we identify in our analysis are often used as tracers of the gas-phase metallicity \citep[e.g.][]{kewley_using_2002,kobulnicky_metallicities_2004,nagao_gas_2006,kewley_metallicity_2008}, which have been shown to exhibit complex relationships with star formation \citep[e.g.][]{mannucci_fundamental_2010,catalan-torrecilla_star_2015}. Like the sSFR, the gas-phase metallicity has also been shown to correlate with the Hubble residual \citep{dandrea_spectroscopic_2011, pan_host_2014}.\par

We have demonstrated the value of an [\ion{O}{II}] equivalent width correction to SN Ia Hubble residuals, however not all SN Ia host galaxies are star forming, so these galaxies will have negligible amounts of [\ion{O}{II}], with the exception of LINER galaxies where the [\ion{O}{II}] emission comes from an AGN \citep{yan_origin_2006}. While Figure \ref{fig:OII_vs_mu} shows that SNe Ia with no detectable EW[\ion{O}{II}] from their host galaxy largely agree with the fitted correlation for star-forming galaxies, there is a smaller population of passive hosts that produce SNe with positive Hubble residuals. More work is needed to understand why this population of SNe are disjoint from the fitted correlation with EW[\ion{O}{II}].

\begin{table}
    \centering
    \rotatebox{90}{
    \begin{minipage}{\textheight}
    \centering
    \caption{The fit parameters and the Pearson and Spearman correlation coefficients for the host stellar mass and the Hubble residual, for different stages of the light-curve correction. We list the slope and significance of the best-fitting linear relationships, and the Pearson and Spearman correlation coefficients for our sample after applying the $x_1$, $c$, Mass ($\delta_\mathrm{host} + \delta_\mathrm{bias}$) corrections and our EW [\ion{O}{II}] light-curve correction. We also include the correlation statistics for the red and blue subsamples for each stage of light-curve correction. The sample contains 37 SNe Ia (19 blue and 18 red) that have a measurable EW[\ion{O}{II}] from their host spectrum.}
    \begin{tabular}{c  c c c c}
    \hline
    Corrections & Slope & Significance $(\sigma)$ & Pearson $r$ & Spearman $r$\\
    \hline
    $x_1 + c$ & $-0.089\pm 0.057$ \textcolor{blue}{($-0.058\pm0.071$)} \textcolor{red}{($-0.113\pm 0.107$)} & $1.55$ \textcolor{blue}{($0.81$)} \textcolor{red}{($1.05$)} & $-0.46 \pm 0.14$ \textcolor{blue}{($-0.39 \pm 0.22$)} \textcolor{red}{($-0.49 \pm 0.19$)} & $-0.43 \pm 0.14$ \textcolor{blue}{($-0.30 \pm 0.20$)} \textcolor{red}{($-0.47 \pm 0.20$)} \\
    $x_1 + c + \mathrm{Mass}$ & $-0.043\pm 0.072$ \textcolor{blue}{($-0.044\pm 0.087$)} \textcolor{red}{($-0.040\pm 0.125$)} & $0.69$ \textcolor{blue}{($0.51$)} \textcolor{red}{($0.32$)} & $-0.29 \pm 0.15$ \textcolor{blue}{($-0.35 \pm 0.22$)} \textcolor{red}{($-0.14 \pm 0.20$)} & $-0.23 \pm 0.14$ \textcolor{blue}{($-0.21 \pm 0.20$)} \textcolor{red}{($-0.12 \pm 0.21$)}\\
    $x_1 + c + [\ion{O}{II}]$ & $-0.022\pm 0.066$ \textcolor{blue}{($0.007\pm 0.070$)} \textcolor{red}{($-0.034\pm 0.135$)} & $0.33$ \textcolor{blue}{($0.09$)} \textcolor{red}{($0.26$)} & $-0.17 \pm 0.16$ \textcolor{blue}{($-0.13 \pm 0.24$)} \textcolor{red}{($-0.08 \pm 0.21$)} & $-0.18 \pm 0.15$ \textcolor{blue}{($-0.20 \pm 0.20$)} \textcolor{red}{($-0.08 \pm 0.22$)} \\
    $x_1 + c + [\ion{O}{II}] + \mathrm{Mass}$ & $-0.006\pm 0.065$ \textcolor{blue}{($-0.009\pm 0.088$)} \textcolor{red}{($0.018\pm 0.125$)} & $0.09$ \textcolor{blue}{($0.11$)} \textcolor{red}{($0.14$)} & $0.14 \pm 0.16$ \textcolor{blue}{($0.00 \pm 0.24$)} \textcolor{red}{($0.34 \pm 0.19$)} & $0.09 \pm 0.14$ \textcolor{blue}{($-0.08 \pm 0.20$)} \textcolor{red}{($0.30 \pm 0.20$)}\\
    \hline
    \end{tabular}
    \end{minipage}}
    \label{tab:mass_OII_corr}
\end{table}

\section{Conclusions} \label{sec:Conclusion}

The well-established `host galaxy mass-step' is a common correction applied to SNe Ia distance moduli, however its physical origin has eluded us since it's discovery. Previous studies have identified correlations between the Hubble residual and a number of host galaxy properties, however many have utilised broad-band photometry to derive host properties, while others have stacked host galaxy spectra together to obtain a high enough S/N for host property measurements.\par

We have completed a spectroscopic study of 75 SNe Ia host galaxies from the Foundation dataset, searching for trends between Hubble residuals and the spectroscopic properties of individual host galaxies. We find evidence of correlations between the Hubble residual and the equivalent widths of gaseous emission features, including \ion{H}{$\alpha$} and [\ion{O}{II}], which are indicators of sSFR. \par

After applying the bias and mass-step corrections, we find that the sample of SNe Ia still produce a correlation between Hubble residual and EW [\ion{O}{II}], a tracer of the sSFR. A correlation also persists between the Balmer decrement and the Hubble residual of blue SNe Ia. This suggests that the bias and mass-step corrections are not fully accounting for the overly simplistic way SN Ia luminosities are standardised.\par

We find that an EW [\ion{O}{II}] correction, with or without a corresponding bias and mass-step correction, is able to correct the host galaxy mass step, reduce the observed relationship between the Balmer decrement and the Hubble residual of blue SNe Ia, and further reduces the scatter in the Hubble Diagram. This relationship between Hubble residual and the EW [\ion{O}{II}] of SN Ia host galaxies, and its relationship with other host galaxy properties, must be explored further if we are to account for all systematics in SN Ia distance measurement, and place tighter constraints on the cosmological parameters that describe our Universe.

\begin{table}
    \centering
    \rotatebox{90}{
    \begin{minipage}{\textheight}
    \centering
    \caption{The fit parameters and the Pearson and Spearman correlation coefficients for the Balmer decrement and the Hubble residual, for different stages of the light-curve correction. We list the slope and significance of the best-fitting linear relationships, and the Pearson and Spearman correlation coefficients for our sample after applying the $x_1$, $c$, Mass ($\delta_\mathrm{host} + \delta_\mathrm{bias}$) corrections and our EW [\ion{O}{II}] light-curve correction. We also include the correlation statistics for the red and blue subsamples for each stage of light-curve correction. The sample contains 30 SNe Ia (15 blue and 15 red) that have both a measurable EW[\ion{O}{II}] and Balmer decrement from their host spectrum.}
    \begin{tabular}{c c c c c}
    \hline
    Corrections & Slope & Significance $(\sigma)$ & Pearson $r$ & Spearman $r$\\
    \hline
    $x_1 + c$ & $0.090 \pm 0.097$ \textcolor{blue}{($0.202\pm 0.120$)} \textcolor{red}{($-0.002\pm 0.136$)} & $0.92$ \textcolor{blue}{($1.68$)} \textcolor{red}{($0.01$)} & $0.20 \pm 0.18$ \textcolor{blue}{($0.62 \pm 0.24$)} \textcolor{red}{($-0.16 \pm 0.26$)} & $0.16 \pm 0.17$ \textcolor{blue}{($0.46 \pm 0.23$)} \textcolor{red}{($-0.10 \pm 0.26$)}\\
    $x_1 + c + \mathrm{Mass}$ & $0.108\pm 0.093$ \textcolor{blue}{($0.190 \pm 0.138$)} \textcolor{red}{($0.017\pm 0.131$)} & $1.20$ \textcolor{blue}{($1.38$)} \textcolor{red}{($0.13$)} & $0.32 \pm 0.17$ \textcolor{blue}{($0.65 \pm 0.23$)} \textcolor{red}{($0.02 \pm 0.26$)} & $0.27 \pm 0.17$ \textcolor{blue}{($0.48 \pm 0.23$)} \textcolor{red}{($0.12 \pm 0.26$)}\\
    $x_1 + c + [\ion{O}{II}]$ & $0.076\pm 0.133$ \textcolor{blue}{($0.085 \pm 0.402$)} \textcolor{red}{($0.044 \pm 0.199$)} & $0.57$ \textcolor{blue}{($0.21$)} \textcolor{red}{($0.22$)} & $0.33 \pm 0.18$ \textcolor{blue}{($0.55 \pm 0.24$)} \textcolor{red}{($0.22 \pm 0.28$)} & $0.30 \pm 0.18$ \textcolor{blue}{($0.47 \pm 0.23$)} \textcolor{red}{($0.20 \pm 0.27$)}\\
    $x_1 + c + [\ion{O}{II}] + \mathrm{Mass}$ & $0.101\pm 0.115$\textcolor{blue}{($0.103\pm 0.739$)} \textcolor{red}{($0.046\pm 0.175$)} & $0.88$ \textcolor{blue}{($0.14$)} \textcolor{red}{($0.26$)} & $0.42 \pm 0.17$ \textcolor{blue}{($0.52 \pm 0.24$)} \textcolor{red}{($0.35 \pm 0.25$)} & $0.37 \pm 0.17$ \textcolor{blue}{($0.45 \pm 0.24$)} \textcolor{red}{($0.26 \pm 0.25$)}\\
    \hline
    \end{tabular}
    \end{minipage}}
    \label{tab:Balmer_OII_corr}
\end{table}

\section*{Acknowledgements}

Based on data acquired at the ANU 2.3-metre Telescope, under program IDs 2360094 and 2370169. The authors acknowledge the traditional custodians of the land on which the ANU 2.3-metre Telescope is based, the Gamilaraay/Kamilaroi people, as well as the traditional custodians of the land on which the ANU Research School of Astronomy and Astrophysics is based, the Ngunnawal and Ngambri peoples. We pay our respects to elders past and present in these communities.\par

The automation of the ANU 2.3-metre telescope was made possible through an initial grant provided by the Centre of Gravitational Astrophysics and the Research School of Astronomy and Astrophysics at the Australian National University and partially through a grant provided by the Australian Research Council through LE230100063.

\section*{Data Availability}

SN Ia light-curve fit parameters and host stellar mass measurements from the Pantheon+ dataset (an amalgamation of surveys including the Foundation Supernova Survey) are available at \url{https://github.com/PantheonPlusSH0ES/DataRelease/}.\par

The spectral properties of the observed Foundation host galaxies used in this analysis can be found at \url{https://doi.org/10.5281/zenodo.13142956}. 
 


\bibliographystyle{mnras}
\bibliography{References}

\begin{thebibliography}{}
\makeatletter
\relax
\def\mn@urlcharsother{\let\do\@makeother \do\$\do\&\do\#\do\^\do\_\do\%\do\~}
\def\mn@doi{\begingroup\mn@urlcharsother \@ifnextchar [ {\mn@doi@} {\mn@doi@[]}}
\def\mn@doi@[#1]#2{\def\@tempa{#1}\ifx\@tempa\@empty \href {http://dx.doi.org/#2} {doi:#2}\else \href {http://dx.doi.org/#2} {#1}\fi \endgroup}
\def\mn@eprint#1#2{\mn@eprint@#1:#2::\@nil}
\def\mn@eprint@arXiv#1{\href {http://arxiv.org/abs/#1} {{\tt arXiv:#1}}}
\def\mn@eprint@dblp#1{\href {http://dblp.uni-trier.de/rec/bibtex/#1.xml} {dblp:#1}}
\def\mn@eprint@#1:#2:#3:#4\@nil{\def\@tempa {#1}\def\@tempb {#2}\def\@tempc {#3}\ifx \@tempc \@empty \let \@tempc \@tempb \let \@tempb \@tempa \fi \ifx \@tempb \@empty \def\@tempb {arXiv}\fi \@ifundefined {mn@eprint@\@tempb}{\@tempb:\@tempc}{\expandafter \expandafter \csname mn@eprint@\@tempb\endcsname \expandafter{\@tempc}}}

\bibitem[\protect\citeauthoryear{Bauer, Drory, Hill  \& Feulner}{Bauer et~al.}{2005}]{bauer_specific_2005}
Bauer A.~E.,  Drory N.,  Hill G.~J.,   Feulner G.,  2005, \mn@doi [ApJ] {10.1086/429289}, 621, L89

\bibitem[\protect\citeauthoryear{Betoule et~al.,}{Betoule et~al.}{2014}]{betoule_improved_2014}
Betoule M.,  et~al., 2014, \mn@doi [\aap] {10.1051/0004-6361/201423413}, 568, A22

\bibitem[\protect\citeauthoryear{Briday et~al.,}{Briday et~al.}{2022}]{briday_accuracy_2022}
Briday M.,  et~al., 2022, \mn@doi [A\&A] {10.1051/0004-6361/202141160}, 657, A22

\bibitem[\protect\citeauthoryear{Brout \& Scolnic}{Brout \& Scolnic}{2021}]{brout_its_2021}
Brout D.,  Scolnic D.,  2021, \mn@doi [ApJ] {10.3847/1538-4357/abd69b}, 909, 26

\bibitem[\protect\citeauthoryear{Brout et~al.,}{Brout et~al.}{2019}]{brout_first_2019}
Brout D.,  et~al., 2019, \mn@doi [ApJ] {10.3847/1538-4357/ab08a0}, 874, 150

\bibitem[\protect\citeauthoryear{Brout et~al.,}{Brout et~al.}{2022a}]{brout_pantheon_2022}
Brout D.,  et~al., 2022a, \mn@doi [ApJ] {10.3847/1538-4357/ac8e04}, 938, 110

\bibitem[\protect\citeauthoryear{Brout et~al.,}{Brout et~al.}{2022b}]{brout_SuperCal_2022}
Brout D.,  et~al., 2022b, \mn@doi [ApJ] {10.3847/1538-4357/ac8bcc}, 938, 111

\bibitem[\protect\citeauthoryear{Campbell, Fraser  \& Gilmore}{Campbell et~al.}{2016}]{campbell_how_2016}
Campbell H.,  Fraser M.,   Gilmore G.,  2016, \mn@doi [MNRAS] {10.1093/mnras/stw115}, 457, 3470

\bibitem[\protect\citeauthoryear{Cappellari \& Emsellem}{Cappellari \& Emsellem}{2004}]{cappellari_parametric_2004}
Cappellari M.,  Emsellem E.,  2004, \mn@doi [PASP] {10.1086/381875}, 116, 138

\bibitem[\protect\citeauthoryear{Catalán-Torrecilla et~al.,}{Catalán-Torrecilla et~al.}{2015}]{catalan-torrecilla_star_2015}
Catalán-Torrecilla C.,  et~al., 2015, \mn@doi [A\&A] {10.1051/0004-6361/201526023}, 584, A87

\bibitem[\protect\citeauthoryear{Childress et~al.,}{Childress et~al.}{2013}]{childress_host_2013}
Childress M.~J.,  et~al., 2013, \mn@doi [ApJ] {10.1088/0004-637X/770/2/108}, 770, 108

\bibitem[\protect\citeauthoryear{D'Andrea et~al.,}{D'Andrea et~al.}{2011}]{dandrea_spectroscopic_2011}
D'Andrea C.~B.,  et~al., 2011, \mn@doi [ApJ] {10.1088/0004-637X/743/2/172}, 743, 172

\bibitem[\protect\citeauthoryear{{DES Collaboration} et~al.,}{{DES Collaboration} et~al.}{2024}]{des_collaboration_dark_2024}
{DES Collaboration} et~al., 2024, The {Dark} {Energy} {Survey}: {Cosmology} {Results} {With} {\textasciitilde}1500 {New} {High}-redshift {Type} {Ia} {Supernovae} {Using} {The} {Full} 5-year {Dataset}, \mn@doi{10.48550/arXiv.2401.02929}, \url {http://arxiv.org/abs/2401.02929}

\bibitem[\protect\citeauthoryear{Dixon et~al.,}{Dixon et~al.}{2022}]{dixon_using_2022}
Dixon M.,  et~al., 2022, \mn@doi [\mnras] {10.1093/mnras/stac2994}, 517, 4291

\bibitem[\protect\citeauthoryear{Domínguez et~al.,}{Domínguez et~al.}{2013}]{dominguez_dust_2013}
Domínguez A.,  et~al., 2013, \mn@doi [ApJ] {10.1088/0004-637X/763/2/145}, 763, 145

\bibitem[\protect\citeauthoryear{Dopita, Hart, McGregor, Oates, Bloxham  \& Jones}{Dopita et~al.}{2007}]{dopita_wide_2007}
Dopita M.,  Hart J.,  McGregor P.,  Oates P.,  Bloxham G.,   Jones D.,  2007, \mn@doi [Ap\&SS] {10.1007/s10509-007-9510-z}, 310, 255

\bibitem[\protect\citeauthoryear{Foley et~al.,}{Foley et~al.}{2018}]{foley_foundation_2018}
Foley R.~J.,  et~al., 2018, \mn@doi [\mnras] {10.1093/mnras/stx3136}, 475, 193

\bibitem[\protect\citeauthoryear{Freedman}{Freedman}{2021}]{freedman_measurements_2021}
Freedman W.~L.,  2021, \mn@doi [ApJ] {10.3847/1538-4357/ac0e95}, 919, 16

\bibitem[\protect\citeauthoryear{Galbany et~al.,}{Galbany et~al.}{2022}]{galbany_aperture-corrected_2022}
Galbany L.,  et~al., 2022, \mn@doi [A\&A] {10.1051/0004-6361/202141568}, 659, A89

\bibitem[\protect\citeauthoryear{Guy et~al.,}{Guy et~al.}{2007}]{guy_salt2_2007}
Guy J.,  et~al., 2007, \mn@doi [\aap] {10.1051/0004-6361:20066930}, 466, 11

\bibitem[\protect\citeauthoryear{Guy et~al.,}{Guy et~al.}{2010}]{guy_supernova_2010}
Guy J.,  et~al., 2010, \mn@doi [\aap] {10.1051/0004-6361/201014468}, 523, A7

\bibitem[\protect\citeauthoryear{Hicken, Challis, Jha, Kirshner, Matheson, Modjaz, Rest  \& Wood-Vasey}{Hicken et~al.}{2009}]{hicken_cfa3_2009}
Hicken M.,  Challis P.,  Jha S.,  Kirshner R.~P.,  Matheson T.,  Modjaz M.,  Rest A.,   Wood-Vasey W.~M.,  2009, \mn@doi [ApJ] {10.1088/0004-637X/700/1/331}, 700, 331

\bibitem[\protect\citeauthoryear{Hicken et~al.,}{Hicken et~al.}{2012}]{hicken_cfa4_2012}
Hicken M.,  et~al., 2012, \mn@doi [ApJS] {10.1088/0067-0049/200/2/12}, 200, 12

\bibitem[\protect\citeauthoryear{Jones et~al.,}{Jones et~al.}{2019}]{jones_foundation_2019}
Jones D.~O.,  et~al., 2019, \mn@doi [ApJ] {10.3847/1538-4357/ab2bec}, 881, 19

\bibitem[\protect\citeauthoryear{Kelly}{Kelly}{2007}]{kelly_aspects_2007}
Kelly B.~C.,  2007, \mn@doi [ApJ] {10.1086/519947}, 665, 1489

\bibitem[\protect\citeauthoryear{Kelly, Hicken, Burke, Mandel  \& Kirshner}{Kelly et~al.}{2010}]{kelly_hubble_2010}
Kelly P.~L.,  Hicken M.,  Burke D.~L.,  Mandel K.~S.,   Kirshner R.~P.,  2010, \mn@doi [ApJ] {10.1088/0004-637X/715/2/743}, 715, 743

\bibitem[\protect\citeauthoryear{Kelsey et~al.,}{Kelsey et~al.}{2021}]{kelsey_effect_2021}
Kelsey L.,  et~al., 2021, \mn@doi [MNRAS] {10.1093/mnras/staa3924}, 501, 4861

\bibitem[\protect\citeauthoryear{{Kenworthy} et~al.,}{{Kenworthy} et~al.}{2021}]{Kenworthy_salt3_2021}
{Kenworthy} W.~D.,  et~al., 2021, \mn@doi [\apj] {10.3847/1538-4357/ac30d8}, \href {https://ui.adsabs.harvard.edu/abs/2021ApJ...923..265K} {923, 265}

\bibitem[\protect\citeauthoryear{Kewley \& Dopita}{Kewley \& Dopita}{2002}]{kewley_using_2002}
Kewley L.~J.,  Dopita M.~A.,  2002, \mn@doi [A\&AS] {10.1086/341326}, 142, 35

\bibitem[\protect\citeauthoryear{Kewley \& Ellison}{Kewley \& Ellison}{2008}]{kewley_metallicity_2008}
Kewley L.~J.,  Ellison S.~L.,  2008, \mn@doi [ApJ] {10.1086/587500}, 681, 1183

\bibitem[\protect\citeauthoryear{Khetan et~al.,}{Khetan et~al.}{2021}]{khetan_new_2021}
Khetan N.,  et~al., 2021, \mn@doi [\aap] {10.1051/0004-6361/202039196}, 647, A72

\bibitem[\protect\citeauthoryear{Kobulnicky \& Kewley}{Kobulnicky \& Kewley}{2004}]{kobulnicky_metallicities_2004}
Kobulnicky H.~A.,  Kewley L.~J.,  2004, \mn@doi [ApJ] {10.1086/425299}, 617, 240

\bibitem[\protect\citeauthoryear{Lampeitl et~al.,}{Lampeitl et~al.}{2010}]{lampeitl_effect_2010}
Lampeitl H.,  et~al., 2010, \mn@doi [ApJ] {10.1088/0004-637X/722/1/566}, 722, 566

\bibitem[\protect\citeauthoryear{Mannucci, Cresci, Maiolino, Marconi  \& Gnerucci}{Mannucci et~al.}{2010}]{mannucci_fundamental_2010}
Mannucci F.,  Cresci G.,  Maiolino R.,  Marconi A.,   Gnerucci A.,  2010, \mn@doi [MNRAS] {10.1111/j.1365-2966.2010.17291.x}, 408, 2115

\bibitem[\protect\citeauthoryear{Meldorf et~al.,}{Meldorf et~al.}{2022}]{meldorf_dark_2022}
Meldorf C.,  et~al., 2022, \mn@doi [MNRAS] {10.1093/mnras/stac3056}, 518, 1985

\bibitem[\protect\citeauthoryear{Nagao, Maiolino  \& Marconi}{Nagao et~al.}{2006}]{nagao_gas_2006}
Nagao T.,  Maiolino R.,   Marconi A.,  2006, \mn@doi [A\&A] {10.1051/0004-6361:20065216}, 459, 85

\bibitem[\protect\citeauthoryear{{Osterbrock}}{{Osterbrock}}{1989}]{osterbrock_astrophysics_1989}
{Osterbrock} D.~E.,  1989, {Astrophysics of gaseous nebulae and active galactic nuclei}.
University Science Books

\bibitem[\protect\citeauthoryear{Pan et~al.,}{Pan et~al.}{2014}]{pan_host_2014}
Pan Y.~C.,  et~al., 2014, \mn@doi [MNRAS] {10.1093/mnras/stt2287}, 438, 1391

\bibitem[\protect\citeauthoryear{Perlmutter et~al.,}{Perlmutter et~al.}{1999}]{perlmutter_measurements_1999}
Perlmutter S.,  et~al., 1999, \mn@doi [ApJ] {10.1086/307221}, 517, 565

\bibitem[\protect\citeauthoryear{Phillips}{Phillips}{1993}]{phillips_absolute_1993}
Phillips M.~M.,  1993, \mn@doi [ApJ] {10.1086/186970}, 413, L105

\bibitem[\protect\citeauthoryear{{Popovic}, {Brout}, {Kessler}  \& {Scolnic}}{{Popovic} et~al.}{2023}]{popovic_pantheon_2022}
{Popovic} B.,  {Brout} D.,  {Kessler} R.,   {Scolnic} D.,  2023, \mn@doi [\apj] {10.3847/1538-4357/aca273}, \href {https://ui.adsabs.harvard.edu/abs/2023ApJ...945...84P} {945, 84}

\bibitem[\protect\citeauthoryear{Rest et~al.,}{Rest et~al.}{2014}]{rest_cosmological_2014}
Rest A.,  et~al., 2014, \mn@doi [ApJ] {10.1088/0004-637X/795/1/44}, 795, 44

\bibitem[\protect\citeauthoryear{Riess et~al.,}{Riess et~al.}{1998}]{riess_observational_1998}
Riess A.~G.,  et~al., 1998, \mn@doi [AJ] {10.1086/300499}, 116, 1009

\bibitem[\protect\citeauthoryear{Riess et~al.,}{Riess et~al.}{2022}]{riess_comprehensive_2022}
Riess A.~G.,  et~al., 2022, \mn@doi [ApJL] {10.3847/2041-8213/ac5c5b}, 934, L7

\bibitem[\protect\citeauthoryear{Rigault et~al.,}{Rigault et~al.}{2013}]{rigault_evidence_2013}
Rigault M.,  et~al., 2013, \mn@doi [\aap] {10.1051/0004-6361/201322104}, 560, A66

\bibitem[\protect\citeauthoryear{Rigault et~al.,}{Rigault et~al.}{2020}]{rigault_strong_2020}
Rigault M.,  et~al., 2020, \mn@doi [\aap] {10.1051/0004-6361/201730404}, 644, A176

\bibitem[\protect\citeauthoryear{Rose, Garnavich  \& Berg}{Rose et~al.}{2019}]{rose_think_2019}
Rose B.~M.,  Garnavich P.~M.,   Berg M.~A.,  2019, \mn@doi [ApJ] {10.3847/1538-4357/ab0704}, 874, 32

\bibitem[\protect\citeauthoryear{Rubin et~al.,}{Rubin et~al.}{2023}]{rubin_union_2023}
Rubin D.,  et~al., 2023, Union {Through} {UNITY}: {Cosmology} with 2,000 {SNe} {Using} a {Unified} {Bayesian} {Framework}, \url {http://arxiv.org/abs/2311.12098}

\bibitem[\protect\citeauthoryear{Sako et~al.,}{Sako et~al.}{2011}]{sako_photometric_2011}
Sako M.,  et~al., 2011, \mn@doi [ApJ] {10.1088/0004-637X/738/2/162}, 738, 162

\bibitem[\protect\citeauthoryear{Salim, Boquien  \& Lee}{Salim et~al.}{2018}]{salim_dust_2018}
Salim S.,  Boquien M.,   Lee J.~C.,  2018, \mn@doi [ApJ] {10.3847/1538-4357/aabf3c}, 859, 11

\bibitem[\protect\citeauthoryear{Salpeter}{Salpeter}{1955}]{salpeter_luminosity_1955}
Salpeter E.~E.,  1955, \mn@doi [ApJ] {10.1086/145971}, 121, 161

\bibitem[\protect\citeauthoryear{Sanchez-Blazquez et~al.,}{Sanchez-Blazquez et~al.}{2006}]{sanchez-blazquez_medium-resolution_2006}
Sanchez-Blazquez P.,  et~al., 2006, \mn@doi [\mnras] {10.1111/j.1365-2966.2006.10699.x}, 371, 703

\bibitem[\protect\citeauthoryear{Scolnic et~al.,}{Scolnic et~al.}{2018}]{scolnic_complete_2018}
Scolnic D.~M.,  et~al., 2018, \mn@doi [ApJ] {10.3847/1538-4357/aab9bb}, 859, 101

\bibitem[\protect\citeauthoryear{Smith et~al.,}{Smith et~al.}{2020}]{smith_first_2020}
Smith M.,  et~al., 2020, \mn@doi [MNRAS] {10.1093/mnras/staa946}, 494, 4426

\bibitem[\protect\citeauthoryear{Stritzinger et~al.,}{Stritzinger et~al.}{2010}]{stritzinger_distance_2010}
Stritzinger M.,  et~al., 2010, \mn@doi [AJ] {10.1088/0004-6256/140/6/2036}, 140, 2036

\bibitem[\protect\citeauthoryear{Sullivan et~al.,}{Sullivan et~al.}{2010}]{sullivan_dependence_2010}
Sullivan M.,  et~al., 2010, \mn@doi [MNRAS] {10.1111/j.1365-2966.2010.16731.x}, 406, 782

\bibitem[\protect\citeauthoryear{Tripp}{Tripp}{1998}]{tripp_two-parameter_1998}
Tripp R.,  1998, \aap, 331, 815

\bibitem[\protect\citeauthoryear{Uddin, Mould, Lidman, Ruhlmann-Kleider  \& Hardin}{Uddin et~al.}{2017}]{uddin_cosmological_2017}
Uddin S.~A.,  Mould J.,  Lidman C.,  Ruhlmann-Kleider V.,   Hardin D.,  2017, \mn@doi [Publ. Astron. Soc. Australia] {10.1017/pasa.2017.2}, 34, e009

\bibitem[\protect\citeauthoryear{Vazdekis, Koleva, Ricciardelli, Röck  \& Falcón-Barroso}{Vazdekis et~al.}{2016}]{vazdekis_uv-extended_2016}
Vazdekis A.,  Koleva M.,  Ricciardelli E.,  Röck B.,   Falcón-Barroso J.,  2016, \mn@doi [MNRAS] {10.1093/mnras/stw2231}, 463, 3409

\bibitem[\protect\citeauthoryear{Vincenzi et~al.,}{Vincenzi et~al.}{2024}]{vincenzi2024dark}
Vincenzi M.,  et~al., 2024, The Dark Energy Survey Supernova Program: Cosmological Analysis and Systematic Uncertainties (\mn@eprint {arXiv} {2401.02945})

\bibitem[\protect\citeauthoryear{Yan, Newman, Faber, Konidaris, Koo  \& Davis}{Yan et~al.}{2006}]{yan_origin_2006}
Yan R.,  Newman J.~A.,  Faber S.~M.,  Konidaris N.,  Koo D.,   Davis M.,  2006, \mn@doi [ApJ] {10.1086/505629}, 648, 281

\makeatother
\end{thebibliography}







\bsp	
\label{lastpage}
\end{document}